\documentclass[%
 aip,
 jcp,%
 amsmath,amssymb,
reprint,%
twocolumn,
]{revtex4-1}

\usepackage{graphicx}
\usepackage{dcolumn}
\usepackage{bm}
\usepackage{subfig}
\usepackage{color}

\begin{document}

\preprint{AIP/123-QED}

\title[Phase separation dynamics in a two-dimensional magnetic mixture]{Phase separation dynamics in a two-dimensional magnetic mixture}

\author{K. Lichtner}
\affiliation{%
Institute of Theoretical Physics, Secr. EW~7-1, Technical University Berlin, \\Hardenbergstr. 36, D-10623 Berlin, Germany
}%
\email{lichtner@mailbox.tu-berlin.de}
\author{A. J. Archer}%
\affiliation{ Department of Mathematical Sciences, Loughborough University, \\Leicestershire, LE11 3TU, UK
}%

\author{S. H. L. Klapp}
\affiliation{%
Institute of Theoretical Physics, Secr. EW~7-1, Technical University Berlin, \\Hardenbergstr. 36, D-10623 Berlin, Germany
}%

\date{\today}

\begin{abstract}
%
Based on classical density functional theory (DFT), we investigate the demixing phase transition of a two-dimensional, binary Heisenberg fluid mixture. The particles in the mixture are modeled as Gaussian soft spheres, where one component is characterized by an additional classical spin-spin interaction of Heisenberg type. Within the DFT we treat the particle interactions using a mean-field approximation. For certain magnetic coupling strengths we calculate phase diagrams in the density-concentration plane. For sufficiently large coupling strengths and densities, we find a demixing phase transition driven by the ferromagnetic interactions of the magnetic species. We also provide a microscopic description (i.e., density profiles) of the resulting non-magnetic/magnetic fluid-fluid interface. Finally, we investigate the phase separation using dynamical density functional theory (DDFT), considering both nucleation processes and spinodal demixing. 
\end{abstract}

\pacs{Valid PACS appear here}
\maketitle

%
\section{\label{sec:Introduction}Introduction}
The theoretical description of the phase separation of fluid mixtures is a long-standing problem with importance in many areas of soft matter physics such as, e.g., the stability of molecular and colloidal solutions\cite{Rowlinsonbook}, the interactions between nanoparticles and macromolecules including novel phenomena such as Casimir forces \cite{Hertlein07}, as well as
interfacial and confinement effects occurring in the presence of surfaces.
Indeed, even ``simple" mixtures consisting of spherical particles can display non-trivial phase behaviour including triple points and critical end points not present for one-component systems 
\cite{Kon,Wilding98}. Correspondingly, even more complex behaviour is observed for particles with internal degrees of freedom such as magnetic particles \cite{Rungsa,Range} and mixtures involving shape-anisotropic particles such as colloidal rods \cite{Dennison11}.

Besides purely numerical approaches such as Monte-Carlo or Molecular (Brownian) dynamics computer simulations, 
classical density functional theory (DFT) has proved to be a very accurate tool for describing both the homogeneous phase behaviour of mixtures and,
at least for simple models, also the inhomogeneous structure occurring at interfaces. The key quantity in DFT is the one-particle density, which is obtained through minimization
of a grand canonical free energy functional corresponding to the microscopic Hamiltonian of the system \cite{evansDFT}.
In addition to yielding the equilibrium phase diagram and microscopic fluid structure, DFT techniques have been successfully used to calculate nucleation barriers for state points in the metastable region of the phase diagrams \cite{oxtoby:7521,talanquer:5190,UC07,ArcherEvansNucleation}. Based on the success of these approaches, it seems very tempting to use DFT techniques also to tackle the {\em non-equilibrium dynamics} of the phase separation, including the growth of nuclei, the actual nucleation pathway and coarsening processes during spinodal decomposition. Indeed, one motivation for studying the phase separation dynamics accompanying demixing transitions is their important role in the context of pattern formation and self-organization \cite{Cross,Kapral}. Traditionally, phase separation dynamics is studied using mesoscopic models involving equations of motion for coarse-grained order parameters
\cite{Hohenberg}. The advantage of addressing these topics using DFT is that the latter allows one to establish the link between the macroscopic behaviour of the system to the
microscopic Hamiltonian, which is naturally incorporated via the excess Helmholtz free energy functional. In the last few years, the 
first steps in these directions have already been made on the basis of the so-called dynamical density functional theory\cite{marconi:8032,marconi:a413, ArcherSpinDec, AR04} (DDFT), which consists of a generalized continuity equation for the one-body density distribution of a
many-particle systems of overdamped (Brownian) colloidal particles. Recent applications of the DDFT to phase separation kinetics include spinodal decomposition in spherical fluids \cite{ArcherSpinDec} and heterogeneous nucleation at solid surfaces \cite{Kahl,Vossen11,PhysRevLett.100.108302}. However, most of these studies have been devoted to simple
fluids with no internal degrees of freedom.

In the present work, we use both static and dynamic DFT to explore
the phase separation of one of the simplest examples
of a mixtures with internal degrees of freedom, that is, 
a binary fluid of spherical particles where one species
carries a classical, (3D) Heisenberg spin. Heisenberg fluids \cite{PhysRevE.58.3426,PhysRevE.49.5169,PhysRevE.58.3478,PhysRevE.52.1915,PhysRevE.55.7242} are basic models for continuum systems exhibiting ferromagnetic order, particularly for the description of ferromagnetism in undercooled liquid metal alloys. Mixtures of such systems and, in particular, mixtures
of magnetic and nonmagnetic particles are promising candidates for the controlled fabrication
of patterns on the micron scale \cite{Rungsa}. 
Moreover, an obvious attractive feature of these systems is that the phase separation and thus, the occurrence of patterns can be tuned by external magnetic fields. 

From the theoretical side, the equilibrium properties of 
one-component Heisenberg fluids \cite{PhysRevE.58.3426,PhysRevE.49.5169,PhysRevE.58.3478,PhysRevE.52.1915,PhysRevE.55.7242},
as well as other spin fluids with two-dimensional (XY) and Ising spins and mixtures thereof have been extensively studied by MC simulations, integral equation methods, and (mean field) density functional theories
(see Ref.\ \onlinecite{Omelyan} and references therein). However, this microscopic level of description for the dynamic behaviour is essentially unexplored.
As a starting point to fill this gap we consider here a Heisenberg mixture in two spatial dimensions in the absence of an external field. The restriction to a 2D situation is actually close to many experiments 
(see, e.g., Ref.~\onlinecite{Rungsa}) and has the advantage that the calculated structure can be easily visualized. Within the vast parameter space characterizing our model, we focus on a situation
where the system demixes into a non-magnetic and a ferromagnetic phase. For this situation, we first use conventional (static) DFT to 
calculate a complete phase diagram (involving a first-order transition and a tricritical point), as well as the inhomogeneous fluid density and magnetization profiles characterizing the liquid interface.
Based on this information we then consider the phase separation dynamics, focussing on the nucleation of non-magnetic bubbles within the ferromagnetic liquid phase. For this problem, we compare 
the results of three different approaches, namely classical nucleation theory (CNT), 
which is based on macroscopic concepts, an approach based on equilibrium DFT, and finally DDFT. We demonstrate that both DFT approaches yield
consistent results for the nucleation barriers, but predict different pathways due to the fact that the DDFT conserves the densities (contrary to DFT). Moreover, the DFT nucleation barriers
differ from the CNT predictions when the size of the critical nucleus becomes small.  We also present evidence that the DDFT can describe the coarsening process during spinodal
decomposition. The remainder of this paper is organized as followed. In Sec.\ II we formulate the model Hamiltonian for the binary system. The equilibrium theory of the demixing transition is presented in Sec.\ III, which includes a calculation of the phase diagram and the interfacial structure. 
In Sec.\ IV we turn to discuss the dynamics of the demixing transition. Finally, we summarize the results in Sec.\ V.\\\\



\section{\label{sec:Model}Model}

The fluid system that we investigate is a binary mixture composed of two species. One species ($A$) is composed of spherical particles which interact via purely isotropic and repulsive forces. The other species ($B$) consists of magnetic particles. In addition to the repulsive interaction induced by the particle cores, these particles carry magnetic moments. The interaction part of the Hamiltonian may therefore be decomposed into a core part and a contribution from the spin-spin interaction\begin{align}
\mathcal{H}^{\text{int}}=\dfrac12 \sum\limits_{\alpha,\beta}\sum\limits_{i,j=1\atop i\neq j}^NV^{\alpha\beta}(\mathbf{r}_i,\mathbf{r}_j,\mathbf{s}_i,\mathbf{s}_j),
\end{align}
where $\alpha,\beta=\{A,B\}$ and
\begin{align}
V^{\alpha\beta}(\mathbf{r}_i,\mathbf{r}_j,\mathbf{s}_i,\mathbf{s}_j)\negthinspace=\negthinspace V_{\mathrm{core}}(\mathbf{r}_i,\mathbf{r}_j)\negthinspace+\negthinspace V_{\text{mag}}(\mathbf{r}_i,\mathbf{r}_j,\mathbf{s}_i,\mathbf{s}_j)\delta_{\alpha,B}\delta_{\beta,B}.\label{eq.1}
\end{align}
In our model, the particles are confined to a (two-dimensional) plane, so that the position of particle $i$, denoted $\mathbf{r}_i=(x_i,z_i)$, but the magnetic moment is represented by a three-dimensional normalized classical spin $\mathbf{s}_i$ whose orientation is described by the Euler angles $\omega=(\theta,\varphi)$. For the magnetic interaction, we choose the Heisenberg model,
\begin{align}
V_{\text{mag}}(|\mathbf{r}-\mathbf{r}'|,\omega,\omega')=&J(|\mathbf{r}-\mathbf{r}'|) \mathbf{s}_1\cdot\mathbf{s}_2\label{eq.3},
\end{align}
where $J(|\mathbf{r}-\mathbf{r}'|)$ determines the range of the spin-spin interaction. We further assume that 
$J(|\mathbf{r}-\mathbf{r}'|)$ can be described by Yukawa's potential, that is,
\begin{align}
J(|\mathbf{r}-\mathbf{r}'|)=\begin{cases}0, &\text{if } |\mathbf{r}-\mathbf{r}'|<\sigma,\\ 
-J\dfrac{e^{-(|\mathbf{r}-\mathbf{r}'|/\sigma-1)}}{|\mathbf{r}-\mathbf{r}'|/\sigma}, &\text{else}.\end{cases}\label{eq.4}
\end{align}
For interparticle distances $|\mathbf{r}-\mathbf{r}'|<\sigma$ the interaction between two magnetic particles is assumed to be small as compared to the repulsion from the core potentials (see below) and we therefore set the Yukawa potential in our model to zero in this region. The sign of the coupling constant $J^*=J/(k_BT)$ in Eq.~(\ref{eq.4}) (where $k_B$ is Boltzmann's constant and $T$ is the temperature) determines which type of spin ordering is preferred. As we show below in Sec.\ \ref{subsect.demixing}, the choice $J^*>0$ yields a spontaneous ferromagnetic ordering of the magnetic component of the mixture at temperatures $T$ below a Curie temperature $T_C$. On the other hand, the choice $J^*<0$ favors antiferromagnetic ordering. To model the repulsion between the particles, we choose a Gaussian with height $\varepsilon$ and width $\sigma$. The resulting ``Gaussian core" model (GCM), first studied by Stillinger~\cite{stillinger:3968}, is given by
\begin{align}
 V_{\mathrm{core}}(|\mathbf{r}-\mathbf{r}'|)=\varepsilon\exp(-|\mathbf{r}-\mathbf{r}'|^2/\sigma^2)\label{eq.2}.
\end{align}
The GCM is often used as an approximation for the effective interactions between the centre of mass of two ``soft'' particles, such as polymers and star-polymers\cite{PhysRevLett.85.2522,Likos2001267} or dendrimers\cite{LSLBPL01,LRD02,GAL06}. The dimensionless quantity $\varepsilon^*=\varepsilon/(k_BT)>0$ determines the strength of the repulsion, and the range parameter $\sigma$ roughly corresponds to the radius of gyration of the `particles'. For the magnetic interaction, we choose a positive coupling constant $J^*>0$. Hence, in the ferromagnetic phase the magnetic contribution to the pair potential acts effectively as an attractive tail to the repulsive (soft) core.\\

\section{Equilibrium theory of the demixing transition}\label{section:EquilibriumTheory}
\subsection{The density functional}\label{section:theDFT}
\noindent
The central quantity in density functional theory is the singlet (one body) density distribution $\rho_\alpha(\mathbf{r},\omega)$. Following other studies of molecular magnetic fluids \cite{PhysRevE.58.3426} we assume that the singlet density can be factorized into a translational (number density) part, $\rho_\alpha(\mathbf{r})$, and an orientational distribution function, $h_\alpha(\mathbf{r},\omega)$, that is,
\begin{align}
\rho_\alpha(\mathbf{r},\omega)=\rho_\alpha(\mathbf{r})h_\alpha(\mathbf{r},\omega).\label{eq.7a}
\end{align} 
The orientational distribution is normalized, so that,
\begin{align}
\negthickspace\int\negthickspace d\omega h_\alpha(\mathbf{r},\omega)=1\label{eq.7b},
\end{align}
which yields $h_A=1/(4\pi)$ for the particles without orientational degrees of freedom (species $A$). The equilibrium fluid singlet density distribution is that which minimizes the grand free energy functional \cite{evansDFT}
\begin{align}
\Omega[\{\rho_\alpha\}]=\mathcal{F}[\{\rho_\alpha\}]-\sum_\alpha\negthickspace\int\negthickspace d\omega\negthickspace\int\negthickspace d\mathbf{r} \Big[\mu_\alpha-V_\text{ext}(\mathbf{r},\omega)\Big]\rho_\alpha(\mathbf{r},\omega),\label{eq.7}
\end{align}
where $\mathcal{F}$ is the Helmholtz free energy functional and $\mu_\alpha$ denotes the chemical potential for species $\alpha$. 
Note also that $\int\negthickspace d\mathbf{r}=\int\negthickspace dx\negthickspace\int\negthickspace dz$ denotes a two-dimensional spatial integral. For a given external potential $V_\text{ext}$ and interaction
potentials, the functional in Eq.~(\ref{eq.7}) has a minimum at the equilibrium density $\rho^0_\alpha(\mathbf{r},\omega)$. The functional $\Omega_V[\{\rho_\alpha^0\}]$ is then identical to the grand canonical potential $\Omega$~(cf. Ref. \onlinecite{evansDFT}). In the present study we set the external potential $V_\text{ext}=0$.
The Helmholtz free energy functional can be split up into two contributions:
\begin{align}
\mathcal{F}[\{\rho_\alpha\}]=&\mathcal{F}_\text{id}[\{\rho_\alpha\}]+\mathcal{F}_\text{ex}[\{\rho_\alpha\}]\label{eq.8},
\end{align}
where $\mathcal{F}_{\mathrm{id}}[\{\rho_\alpha\}]=\sum\limits_\alpha\int d\mathbf{r}\int d\omega \rho_\alpha(\mathbf{r},\omega)[\ln(\Lambda_\alpha^2\rho_\alpha(\mathbf{r},\omega))-1]$ is the ideal gas contribution and $\mathcal{F}_\text{ex}[\{\rho_\alpha\}]$ is the excess part. Using Eqs.\ (\ref{eq.7a}) and (\ref{eq.7b}), the ideal gas contribution becomes
\begin{align}
\mathcal{F}_{\mathrm{id}}[\{\rho_\alpha\}]=&k_BT\negthickspace\int\negthickspace d\mathbf{r}\rho_A(\mathbf{r})\Big[\ln(\Lambda_A^2\rho_A(\mathbf{r}) )-\ln 4\pi-1\Big]\\
&+k_BT\negthickspace\int\negthickspace d\mathbf{r}\rho_B(\mathbf{r})\Big[\ln(\Lambda_B^2\rho_B(\mathbf{r}) )-1\Big]\nonumber\\
&+k_BT\negthickspace\int\negthickspace d\mathbf{r}\rho_B(\mathbf{r})\negthickspace\int\negthickspace d\omega h_B(\mathbf{r},\omega) \ln\Big[h_B(\mathbf{r},\omega)\Big]\label{eq.9},
\end{align}
where $\Lambda_\alpha$ denotes the thermal de Broglie wavelength of species $\alpha$. The particle interactions enter into the excess part of the free energy functional, which can be written as\cite{evansDFT}
\begin{align}
\mathcal{F}_\text{ex}[\{\rho_i\}]=&\frac12\sum_{\alpha,\beta}\int\limits_0^1\negthickspace d\lambda\negthickspace\int\negthickspace d\mathbf{r}\negthickspace \int\negthickspace d\mathbf{r'}\negthickspace\int\negthickspace d\omega\negthickspace\int\negthickspace d\omega' \nonumber\\ 
&\times \rho^{(2)}_{\alpha\beta}(\mathbf{r},\mathbf{r'},\omega,\omega';\lambda)V^{\alpha\beta}(|\mathbf{r}-\mathbf{r'}|,\omega,\omega')\label{eq.10}.
\end{align}
Equation~(\ref{eq.10}) is exact for systems with pair interactions. The function $\rho^{(2)}_{\alpha\beta}(\mathbf{r},\mathbf{r'},\omega,\omega';\lambda)$ is the two-body density distribution function which is determined by the pair interactions $V^{\alpha\beta}(|\mathbf{r}-\mathbf{r'}|,\omega,\omega')$, and $\lambda$ is a ``charging" parameter\cite{evansDFT}. In general, the function $\rho_{\alpha\beta}^{(2)}$ is not known exactly. Here we employ a mean-field (MF) approximation by setting $\rho^{(2)}_{\alpha\beta}(\mathbf{r},\mathbf{r'},\omega,\omega';\lambda)=\rho_\alpha(\mathbf{r},\omega)\rho_\beta(\mathbf{r'},\omega')$, i.e.\ that the pair correlation function is set to one. Previous studies\cite{Likos2001267, PhysRevE.62.7961} have shown that this simple MF approximation for the GCM yields reliable results for the fluid structure and thermodynamics, particularly at higher densities. Within the MF approximation, the excess part of the Helmholtz free energy may be written as the following sum
\begin{align}
\mathcal{F}_\text{ex}&=\mathcal{F}_\text{ex}^{AA}+\mathcal{F}_\text{ex}^{BB}+\mathcal{F}_\text{ex}^{AB}+\mathcal{F}_\text{ex}^{BA}
\end{align}
where the contribution due to the interactions between the non-magnetic particles is
\begin{align}
\mathcal{F}_\text{ex}^{AA}&[\left\{\rho_A\right\}]=\dfrac{1}{2}\int\negthickspace d\mathbf{r} \negthickspace\int\negthickspace d\mathbf{r}'\rho_A(\mathbf{r})\rho_A(\mathbf{r}')V_{\text{core}}(|\mathbf{r}-\mathbf{r}'|)\label{eq.12},
\end{align}\label{eq.10a}
and for the magnetic particles
\begin{align}
\mathcal{F}_\text{ex}^{BB}&[\left\{\rho_B\right\}]=\dfrac{1}{2}\int\negthickspace d\omega\negthickspace\int\negthickspace d\omega' h_B(\mathbf{r},\omega)h_B(\mathbf{r}',\omega')\int\negthickspace d\mathbf{r} \negthickspace\int\negthickspace d\mathbf{r}'\nonumber\\
&\thickspace\thickspace\thickspace\thickspace\times\rho_B(\mathbf{r})\rho_B(\mathbf{r}')\Big[V_{\text{core}}(|\mathbf{r}-\mathbf{r}'|)+J(|\mathbf{r}-\mathbf{r}'|)\mathbf{s}\cdot\mathbf{s'}\Big].\label{eq.13}
\end{align}
The contributions $\mathcal{F}_\text{ex}^{AB}$ and $\mathcal{F}_\text{ex}^{BA}$ to the excess free energy functional are equal because of the symmetry of the pair potentials between species $A$ and $B$ (cf. Eq.~(\ref{eq.1})), that is,
\begin{align}
\mathcal{F}_\text{ex}^{AB}[\left\{\rho_\alpha\right\}]&=\dfrac{1}{2}\int\negthickspace d\mathbf{r} \negthickspace\int\negthickspace d\mathbf{r}'\rho_A(\mathbf{r})\rho_B(\mathbf{r}')V_{\text{core}}(|\mathbf{r}-\mathbf{r}'|)\nonumber\\&=\mathcal{F}_\text{ex}^{BA}[\left\{\rho_\alpha\right\}]
\end{align} 
The equilibrium densities $\rho_\alpha^{(0)}(\mathbf{r},\omega)$ are found by minimizing the grand free energy functional given in Eq.~(\ref{eq.7}):
\begin{align}
&\left. \dfrac{\partial \Omega[\rho_\alpha, h]}{\partial \rho_\alpha(\mathbf{r})}\right|_{\rho_\alpha^{(0)}(\mathbf{r})}=0\nonumber,\\
&\left. \dfrac{\partial \Omega[\rho_\alpha, h]}{\partial h_\alpha(\mathbf{r},\omega)}\right|_{h_\alpha^{(0)}(\mathbf{r},\omega)}=0.\label{eq.14a}
\end{align}
Equation (\ref{eq.14a}) yields an implicit equation for $h_B(\mathbf{r},\omega)$,
\begin{align}
 h_B(\mathbf{r},\omega)=\dfrac{\exp(\mathbf{B}(\mathbf{r})\cdot \mathbf{s}(\omega))}{\int d\omega \exp(\mathbf{B}(\mathbf{r})\cdot \mathbf{s}(\omega))},
\end{align}
where the effective field
\begin{align}
\mathbf{B}(\mathbf{r})= -\negthickspace\int\negthickspace d\mathbf{r}'\negthickspace\int\negthickspace d\omega'\rho_B(\mathbf{r}')h_B(\mathbf{r}',\omega')J(|\mathbf{r}-\mathbf{r}'|)\mathbf{s'}\label{eq.15}.
\end{align}
Thus, the orientational distribution is determined solely by the scalar product between the spin and the effective field $\mathbf{B}(\mathbf{r})$. This is an exact result within the MF approximation, which has been previously applied also to three-dimensional Heisenberg fluids\cite{PhysRevE.52.1915, PhysRevE.55.7242,PhysRevE.58.3426} as well as in other contexts such as in liquid crystal theory.\cite{PhysRevA.38.2022}

In the low-temperature ferromagnetic state, the orientational order is uniaxial with respect to a director $\mathbf{n}$. Thus, the angular distribution reduces to $h_B(\mathbf{r},\omega)=h_B(\mathbf{r},u)$, where $u=\mathbf{s}\cdot\mathbf{n}=\cos\theta$. The integration over orientation in Eq.~(\ref{eq.15}) then becomes\cite{PhysRevE.58.3426,PhysRevE.55.7242}
\begin{align}
\int\negthickspace d\omega' h_B(\mathbf{r}',\omega')\mathbf{s'}\longrightarrow 2\pi\int\limits_{-1}^1 du h_B(\mathbf{r},u)u=L(\mathbf{r}),\label{eq.14}
\end{align}
where $L(\mathbf{r})=\coth B(\mathbf{r})-1/B(\mathbf{r})$ is the Langevin function. The latter also defines the local magnetization $m(\mathbf{r})=\int d\omega h_B(\mathbf{r},\omega)\cos\theta=L(\mathbf{r})$. Inserting Eq.~(\ref{eq.14}) into Eq.~(\ref{eq.13}), the Heisenberg contribution to the free energy functional can be written as
\begin{align}
\mathcal{F}_{\text{ex}}^{BB}[\rho]=&\dfrac{1}{2} \int\negthickspace d\mathbf{r}\negthickspace \int\negthickspace d\mathbf{r}'\rho_B(\mathbf{r})\rho_B(\mathbf{r}')\nonumber\\
&\thickspace\thickspace\thickspace\thickspace\times\Big[V_{\text{core}}(|\mathbf{r}-\mathbf{r}'|)+L(\mathbf{r})L(\mathbf{r}')J(|\mathbf{r}-\mathbf{r}'|)\Big]\label{eq.16}.
\end{align}

\subsection{Phase behavior}\label{subsect.demixing}

In this section, we employ the density functional approach introduced above to investigate the phase behavior of the bulk binary mixture. We restrict ourselves to fluid phases. The state of the system can be characterized by the strength of the repulsion $\varepsilon^*=\varepsilon/(k_BT)$, the magnetic coupling parameter $\delta=J/\varepsilon$, the reduced total number density $\rho\sigma^2$ in the system, and the concentration $x$ of the magnetic component. The densities of the individual component can then be written as $\rho_A=(1-x)\rho$ and $\rho_B=x\rho$. Assuming that the system is homogeneous, the Helmholtz free energy per particle $f=\mathcal{F}/N$ follows from Eqs.~(\ref{eq.8})-(\ref{eq.10}) as
\begin{align}
f^{\text{MF}}&(\rho,x)=f_{\text{id}}(\rho,x)+f_{\text{ex}}^{\text{MF}}(\rho,x)\nonumber\\
=&x\ln x + (1-x)\ln(1-x)+x\int\negthickspace d\omega h_B(\omega) \ln\Big[h_B(\omega)\Big]+\nonumber\\
&\dfrac12\rho\Big[(1-x)^2\hat{V}^{AA}(0)+2x(1-x)\hat{V}^{AB}(0)+x^2 \hat{V}^{BB}(0)\Big]\label{eq.18}.
\end{align}
In Eq.~(\ref{eq.18}), the coupling matrix elements $\hat{V}^{AA}(0)$ and $\hat{V}^{AB}(0)$ denote the integrated strength of the repulsive core potential, or equivalently the $k\rightarrow 0$ limit of the Fourier transform of Eq.~(\ref{eq.2}), that is,
\begin{align}
\hat{V}^{AA}(0)&=\int\negthickspace d\mathbf{r}V_{\mathrm{core}}(|\mathbf{r}|)=\hat{V}^{AB}(0).
\end{align}
The element $\hat{V}^{BB}(0)$ involves, in addition, an integral over the magnetic interaction, i.e.
\begin{align}
\hat{V}^{BB}(0)&=\int\negthickspace d\mathbf{r}V_{\mathrm{core}}(|\mathbf{r}|)\nonumber\\
&\thickspace\thickspace\thickspace\thickspace +\int\negthickspace d\mathbf{r} \negthickspace\int\negthickspace d\omega \negthickspace\int\negthickspace d\omega' h_B(\omega)h_B(\omega')J(|\mathbf{r}|)\mathbf{s}\cdot\mathbf{s'}.
\end{align}
In the following, we investigate the possibility of fluid-fluid phase separation. The thermodynamic stability conditions for a binary mixture are given by \cite{PhysRevE.62.7961, PhysRevE.64.041501}
\begin{align}
 &\left(\dfrac{\partial^2 f}{\partial \nu^2}\right)_x > 0\nonumber,\\
&\left(\dfrac{\partial^2 f}{\partial x^2}\right)_\nu > 0\nonumber,\\
&\left(\dfrac{\partial^2 f}{\partial \nu^2}\right)_x\left(\dfrac{\partial^2 f}{\partial x^2}\right)_\nu-\left(\dfrac{\partial^2 f}{\partial \nu\partial x}\right)^2 > 0\label{eq.24},
\end{align}
where $\nu=1/\rho$ is the volume per particle. The first stability condition expresses that the compressibility must be positive, the second ensures stability against spontaneous demixing at constant volume, and the last inequality is the condition for stability at constant pressure. As shown in previous studies\cite{PhysRevE.64.041501}, it is more convenient to use these stability conditions in a constant-pressure ensemble. To this end, we perform a Legendre transform of the Helmholtz free energy per particle, yielding the Gibbs free energy per particle
\begin{align}
 g(x,P)= f(x,\nu)-\left(\dfrac{\partial f}{\partial \nu}\right)_x\nu\label{eq.25}.
\end{align}
For two phases $\text{I}$ and $\text{II}$ to coexist in equilibrium, the chemical potentials of each species $\alpha=A,B$ have to be equal, and the same holds for the pressure and the temperature. In other words, one has $\mu_\alpha^\text{I}(\rho_A^\text{I},\rho_B^\text{I})=\mu_\alpha^\text{II}(\rho_A^\text{II},\rho_B^\text{II})$, $P^\text{I}(\rho_A^\text{I},\rho_B^\text{I})=P^\text{II}(\rho_A^\text{II},\rho_B^\text{II})$ and $T^\text{I}(\rho_A^\text{I},\rho_B^\text{I})=T^\text{II}(\rho_A^\text{II},\rho_B^\text{II})$, where $\rho_\alpha^\text{I(II)}$ are the densities of the components in the two phases. These equilibrium conditions lead to a common-tangent construction on the Gibbs free energy,
\begin{align}
 \left.\left(\dfrac{\partial g}{\partial x}\right)_P \right|_{x_\text{I}}=\left.\left(\dfrac{\partial g}{\partial x}\right)_P \right|_{x_\text{II}}=\dfrac{g(x_\text{I},P)-g(x_\text{II},P)}{x_\text{I}-x_\text{II}}\label{eq.26},
\end{align}
where $P$ is the (bulk) pressure at coexistence. The spinodal is given by the inflection points of $g$, that is
\begin{align}
\left(\dfrac{\partial^2 g}{\partial x^2}\right)_P = 0\label{eq.27}.
\end{align}

\begin{figure*}[tpb]
\includegraphics[width=8cm]{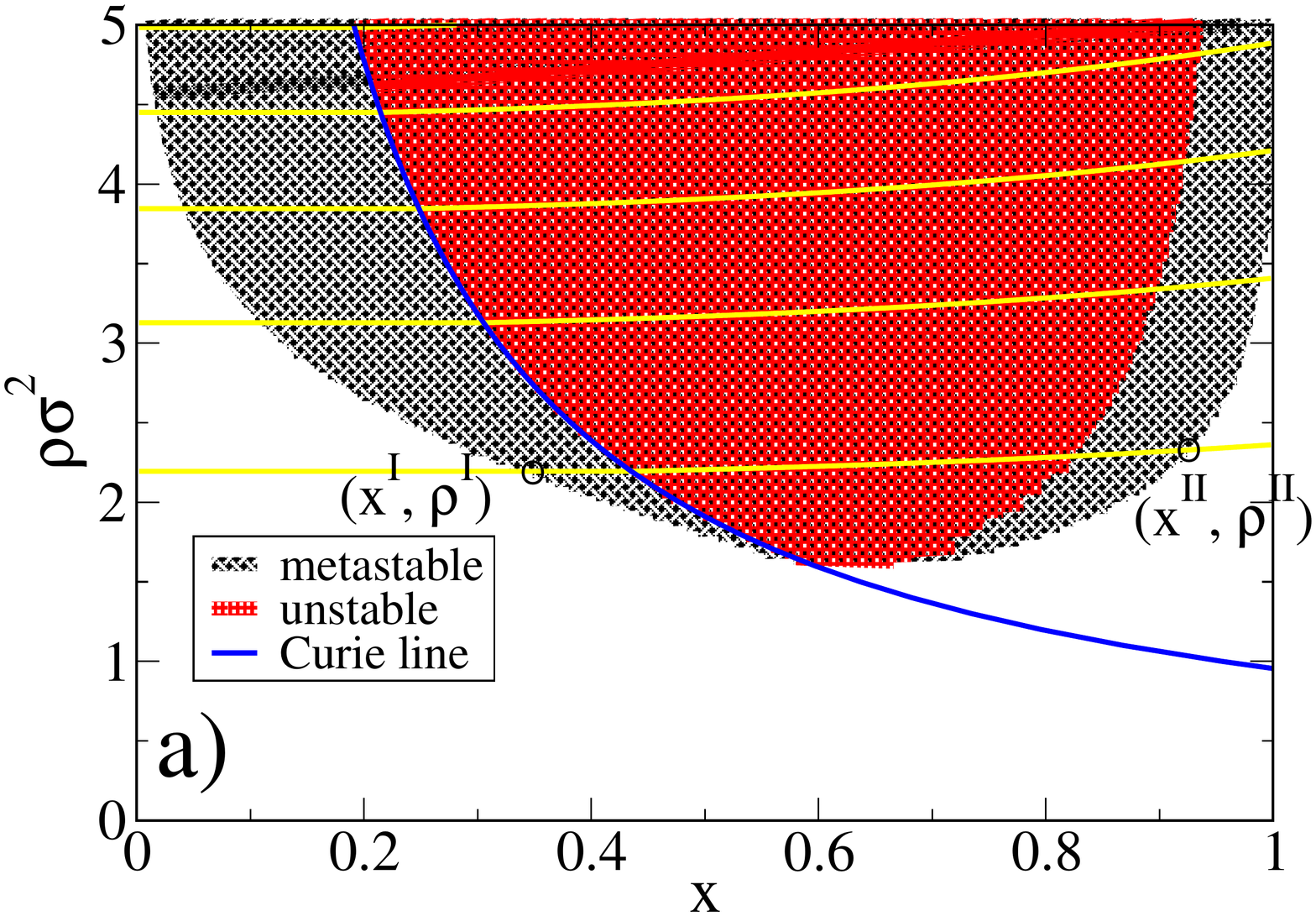}\hspace*{0.05in}
\includegraphics[width=8cm]{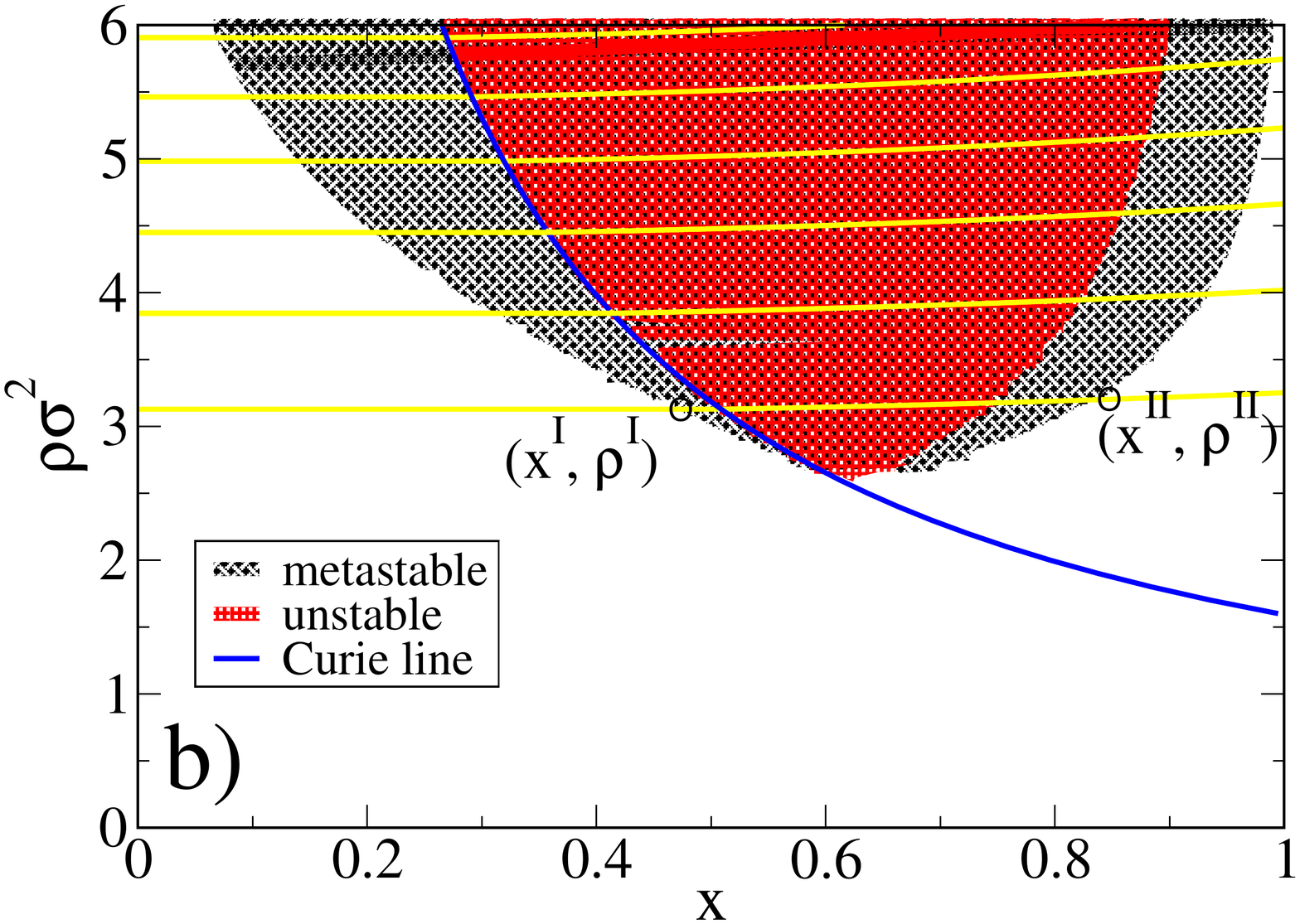}
\caption{The phase diagram for a two-component GCM mixture, where one component is characterized by an additional Heisenberg interaction. The coupling parameters are a) $\varepsilon^*=5.0$, $\delta=0.1$ and b) $\varepsilon^*=5.0$, $\delta=0.06$. The yellow lines are isobars with line-to-line pressure difference of $\Delta P^*=\Delta P\sigma^2/(k_BT)=40$. The black circles denote coexisting state points for $P^*=40$ (a) and $P^*=80$ (b), respectively. In a) the highest isobar indicated corresponds to $P^*=160$, and the (tri)critical point occurs at $\rho_c\sigma^2=1.5$, $x_c=0.63$. The corresponding data in b) are $P^*=280$, $\rho_c\sigma^2=2.6$, and $x_c=0.61$.}
\label{fig.1}
\end{figure*}

Figure \ref{fig.1} shows the phase diagram for the (2D) bulk binary mixture whose free energy is given by Eq.~(\ref{eq.18}). We consider a fixed repulsion strength $\varepsilon^*=5.0$ and two different magnetic coupling parameters, $\delta=J^*/\varepsilon^*=0.1$ and $\delta=0.06$. For both parameter sets we find a first-order demixing phase transition appearing at densities above a critical density $\rho_c$. Moreover, the demixing is coupled to a transition from a paramagnetic phase ($m=0$), which is rich in $A$-particles, to a ferromagnetic phase ($m>0$) rich in $B$-particles. The onset of magnetic order is determined by the Curie line (blue line in Fig. 1), which is obtained by making a Taylor expansion of the Langevin function $L$ [see Eq.~(\ref{eq.14}) and below] around $m=0$ combined with the expression for the effective field $B$ [see Eq.~(\ref{eq.15})] (note that we consider spatially homogeneous systems here). The resulting critical concentration $x_{\text{Curie}}$ as function of the total density is given by
\begin{align}
x_{\text{Curie}}(\rho,J^*)=\dfrac{3}{2\pi J^* \rho \sigma^2}.
\end{align}
Inspecting the position of the Curie line in the phase diagrams in Fig. \ref{fig.1}, we see that the system is entirely disordered, regardless of the concentration, for values of the total density $\rho\sigma^2\lesssim 1$ ($1.5$) for $\delta=0.1$ ($0.06$). Increasing the density from these values towards the critical density, the transition from the paramagnetic into the ferromagnetic phase is of second order, until the Curie line meets with the demixing coexistence curve. This merging occurs directly at the demixing critical density (and critical concentration), corresponding to a {\em tricritical} point. At densities $\rho>\rho_c$, the Curie line then coincides with the low-concentration branch of the demixing spinodal. This reflects the fact that it is the spin-spin interaction [see Eq.~(\ref{eq.1})] which drives the phase separation. Indeed, as can be seen from Eq.~(\ref{eq.16}), the spin-spin interaction reduces the free energy of the system whenever the magnetization is non-zero.

The demixing spinodal has been calculated using Eq.~(\ref{eq.27}). Inside the spinodal the mixture is thermodynamically unstable, as indicated by the red areas in Fig.\ \ref{fig.1}. The black areas in Fig.\ \ref{fig.1} indicate the metastable regions between the spinodal and the coexistence curve. The corresponding coexisting densities $\rho^{\text{I}}$, $\rho^{\text{II}}$ and concentrations $x^\text{I}$, $x^\text{II}$ are calculated using Eq.~(\ref{eq.26}). Recall that coexisting phases are at equal pressure; we display a number of isobars in the density-concentration plane, which are indicated by the yellow lines in Fig. \ref{fig.1}. For each value of the interaction parameter $\delta$ we also display, for one particular exemplary pressure value, a pair of coexisting state points (see circles). From this one may observe that the first-order phase transition is indeed mainly a demixing transition in the sense that the total density change on crossing the transition is small. Finally, a comparison of Figs. 1a) and b) reveals that as $\delta$ is reduced the critical point shifts to larger values of $\rho$. This is because an increase of the number density of the magnetic component supports the ferromagnetic phase transition \big[cf. Eq.~(\ref{eq.15})\big]. 

\subsection{Interfacial structure}
\label{sec:interface_structure}

In Sec. \ref{subsect.demixing} we showed that the magnetic mixture displays a first-order demixing phase transition for a broad range of densities $\rho$ and concentrations $x$. In this section, we focus on the structure of the fluid-fluid interface between the two demixed phases for densities above $\rho_c$. The grand canonical free energy $\Omega$ for the non-uniform binary mixture is given by Eq.~(\ref{eq.7}). Setting the external potential $V_{\text{ext}}=0$ yields a well-defined (one-dimensional) interface between one region enriched with particles from the non-magnetic species and a second region enriched with magnetic particles.  Setting the functional derivative of Eq.~(\ref{eq.7}) to zero yields the Euler-Lagrange equations for the chemical potential of species $A$
\begin{align}
\mu_A=k_BT\ln\Big[\rho_A(z)\Big]+\int\negthickspace d\mathbf{r}'\Big[\rho_A(z')+\rho_B(z')\Big]V_{\text{core}}(|\mathbf{r}-\mathbf{r'}|)\label{eq.mua},
\end{align}
and species $B$, respectively,
\begin{align}
\mu_B&=k_BT\Bigg\{\ln\Big[\rho_B(z)\Big]+B(z)\coth\Big[B(z)\Big]-1\nonumber\\
&+\ln\left[\frac{B(z')}{\sinh\left[B(z')\right]}\right]\Bigg\}\nonumber\\
&+\int\negthickspace d\mathbf{r'}\Big[\rho_A(z')+\rho_B(z')\Big]\beta V_{\text{core}}(|\mathbf{r}-\mathbf{r'}|)\nonumber\\
&+L\Big[B(z)\Big]\int\negthickspace d\mathbf{r'} L\Big[B(z')\Big]\rho_B(z')\beta J(|\mathbf{r}-\mathbf{r'}|)\label{eq.mub},
\end{align}
where $\beta=1/(k_BT)$. 
\begin{figure}[tpb]
\begin{center}
\includegraphics[width=8cm]{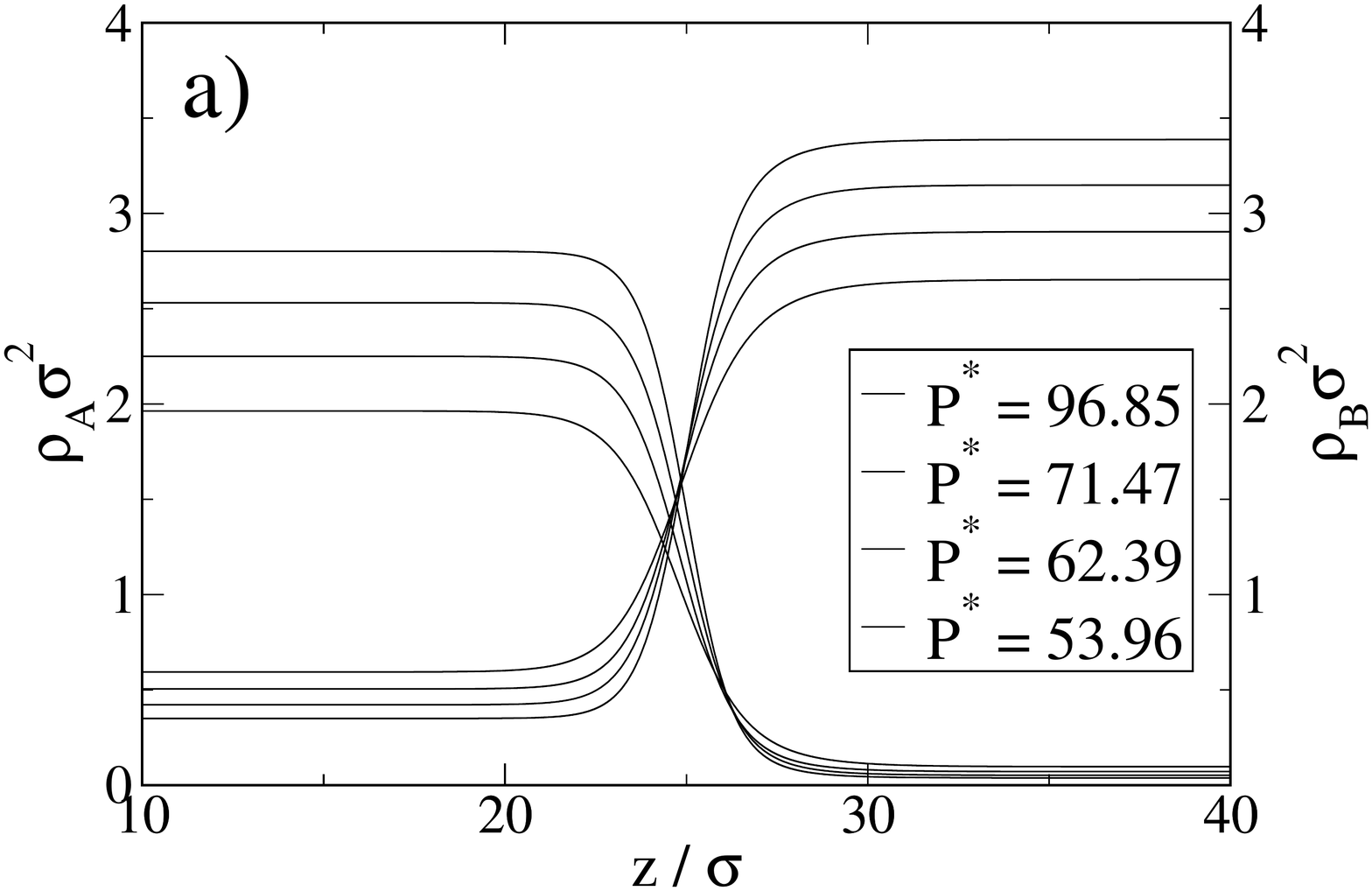}\\
\includegraphics[width=8cm]{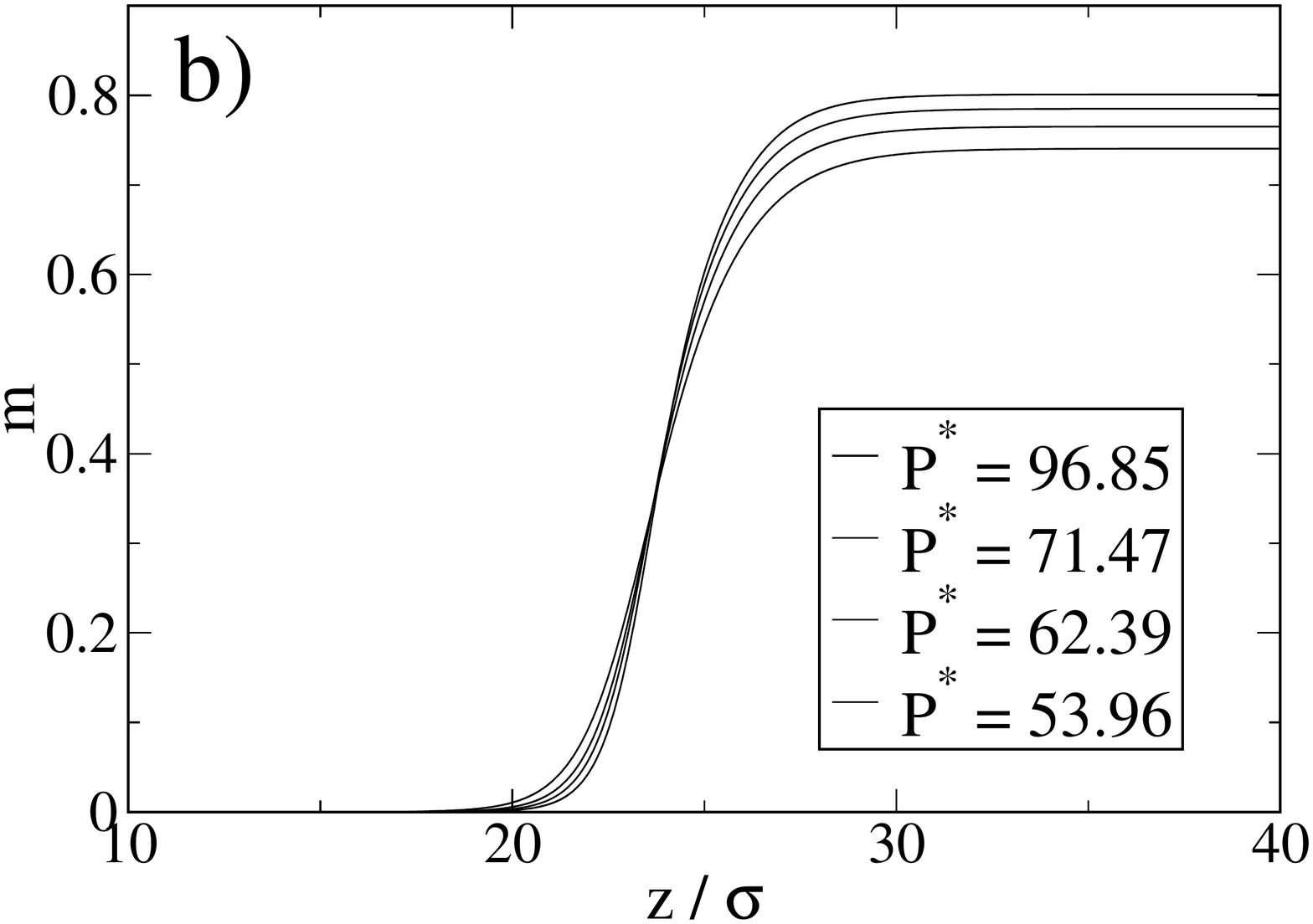}
\end{center}
\caption{a) The density profiles of the two demixed phases and b) the magnetization profile (species $B$) at the interface for different values of $P^*$. The coupling parameters are $\varepsilon^*=5.0$ and $\delta=0.1$.}
\label{fig.2}
\end{figure}
Using the (bulk) coexisting densities $\rho_A^\text{bulk}=\rho_A^\text{I}$, $\rho_B^\text{bulk}=\rho_B^\text{II}$ found from the calculation of the binodal in section \ref{subsect.demixing}, 
the chemical potentials $\mu_A$ and $\mu_B$ can be eliminated from Eqs.~(\ref{eq.mua}) and (\ref{eq.mub}). This leads to the equations for the one-body density profiles across the free interface. Specifically, we obtain for the non-magnetic component
\begin{align}
 &\rho_A(z)=\rho_A^{\text{bulk}}\nonumber\\
&\thickspace\thickspace\thickspace\thickspace\times\exp\left[\sum\limits_{\alpha=A}^B\int\negthickspace d\mathbf{r'}\Big(\rho_\alpha^{\text{bulk}}-\rho_\alpha(z')\Big)\beta V_{\text{core}}(|\mathbf{r}-\mathbf{r'}|)\right],\label{eq.32}
\end{align}
and for the magnetic component
\begin{align}
\rho_B&(z)=\rho_B^{\text{bulk}}\exp\left[\sum\limits_{\alpha=A}^B\int\negthickspace d\mathbf{r'}\Big(\rho_\alpha^{\text{bulk}}-\rho_\alpha(z')\Big)\beta V_{\text{core}}(|\mathbf{r}-\mathbf{r'}|)\right.\nonumber\\
&\left.+L(B^\text{bulk})^2\int\negthickspace d\mathbf{r'}\Big(\rho_B^{\text{bulk}}-\rho_2(z')\Big)\beta J(|\mathbf{r}-\mathbf{r'}|)\right.\nonumber\\
&+\frac{B^\text{bulk}}{\sinh(B^\text{bulk})}-\frac{B(z')}{\sinh[B(z')]}+B^\text{bulk}L^\text{bulk}-B(z')L(z')\nonumber\\
&\left.-L\Big[B(z)\Big]\int\negthickspace d\mathbf{r'}L\Big[B(z')\Big]\Big(\rho_B^{\text{bulk}}-\rho_2(z')\Big)\beta J(|\mathbf{r}-\mathbf{r'}|)\right]\label{eq.33}.
\end{align}
Equations (\ref{eq.32}) and (\ref{eq.33}) can be solved self-consistently. The results for the density profiles and the magnetization are shown in Fig.~\ref{fig.2} for a fixed magnetic coupling parameter $\delta=0.1$ (see Fig. 1a) for the corresponding phase diagram). The decay of the density profiles into the two bulk phases in Fig.\ \ref{fig.2} is monotonic. In previous studies of binary mixtures of soft particles \cite{PhysRevE.64.041501, ALE2002} and also of colloid-polymer mixtures \cite{EvansMolPhys93, Brader02} it was found that non-monotonic oscillatory decay of the density profiles can occur for the free interface between coexisting state points that are sufficiently far removed in the phase diagram from the critical point. We expect a similar scenario for the present system as that observed in the systems studied in Refs.\ \onlinecite{PhysRevE.64.041501, ALE2002}, i.e.\ we expect to observe oscillatory behavior in the density profiles also for the present system, but for higher values of the total bulk density, i.e.\ further away from the critical point. Note that the freezing transition\cite{Likos_2001} of the pure GCM fluid in three dimensions occurs only at much lower temperatures (i.e.\ much larger values of $\varepsilon^*$) than we consider here.

Approaching the critical density from above, the interface softens, resulting in a reduced pressure parallel to the interface. Physically, this softening is reflected by a decreasing line tension (in 2D). To calculate the line tension we assume that the dividing surface is a line that is orientated along the $x$-direction. The pressure tensor $\mathbf{P}$ is a ($2\times2$)-matrix characterized by one component parallel to the interface ($P_T$) and one component normal to the interface ($P_N$), that is
\begin{align}
\mathbf{P}=P_T(z)\hat{e}_x\hat{e}_x+P_N(z)\hat{e}_z\hat{e}_z.
\end{align}
\begin{figure}[tpb]
\begin{center}
\includegraphics[width=8cm]{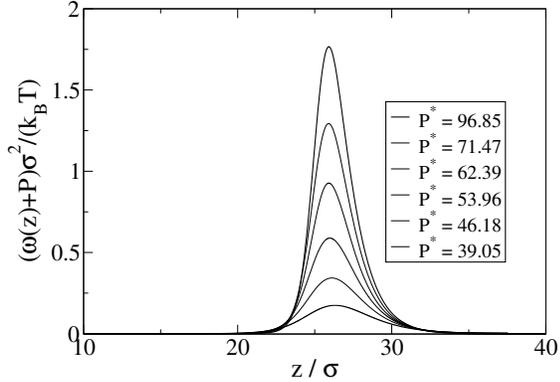}
\end{center}
\caption{The grand potential density as function of $z$ with the interface being located at $z=25.6\sigma$. The coupling parameters are $\varepsilon^*=5.0$ and $\delta=0.1$.}
\label{fig.3}
\end{figure}
Here, $\hat{e}_x$ and $\hat{e}_z$ are normalized unit vectors in the $x$-direction and $z$-direction, respectively. In equilibrium, the normal pressure $P_N(z)$ is constant and equals the bulk pressure $P$ at coexistence. Furthermore, the tangential component $P_T$ only depends on $z$. The interfacial (line) tension $\gamma$ is then defined as the excess force resulting from the dividing interface\cite{Lovett199393,rowlinson}, that is,
\begin{align}
\gamma=\int\limits_{-\infty}^{\infty} dz \left(P-P_T(z)\right).\label{eq.surftens}
\end{align}
To actually calculate $P_T(z)$, we use the relation\cite{Lovett199393} $P_T(z)=-\omega(z)$, where $\omega(z)$ is defined as the grand canonical free energy per unit length 
 calculated in the absence of an external potential. Numerical results for the function $(\omega(z)+P)$ [i.e., the integrand in Eq.~(\ref{eq.surftens})] at various total densities $\rho\sigma^2$ are shown in Fig. 3. We only find non-vanishing values of the function $(\omega(z)+P)$ near the interface. This reflects the simple fact that the interfacial tension stems from the density inhomogeneities at the interface (see Fig. 2). From Fig. 2 we see that the density profiles become smoother as the critical point is approached. This behavior is mirrored by the function $(\omega(z)+P)$ (see Fig. 3), resulting in a vanishing interfacial tension at the critical point. The behavior of $\gamma$ as a function of the pressure difference $(P-P_c)$ is displayed in Fig.~\ref{fig.4}.   

\begin{figure}[tpb]
\begin{center}
\includegraphics[width=8cm]{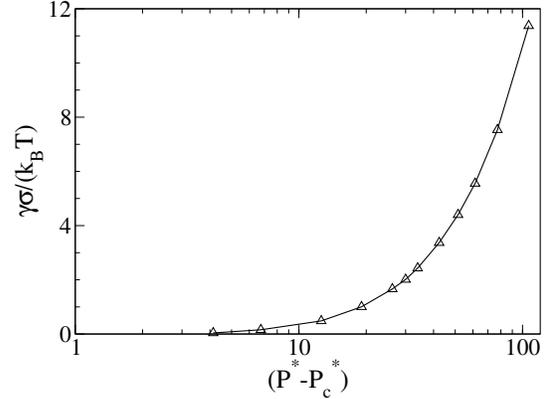} 
\end{center}
\caption{The interfacial (line) tension $\gamma^*=\gamma\sigma/(k_BT)$ between the demixed phases as function of the pressure difference relative to the critical point. The coupling parameters are $\delta=0.1$, $\varepsilon^*=5.0$.}
\label{fig.4}
\end{figure}

\section{\label{sec:Demixing dynamics}Demixing dynamics}

We now turn to discuss the dynamics of phase separation in the present system. To this end, we employ the DDFT approach,\cite{marconi:8032,marconi:a413, ArcherSpinDec, AR04} in which the time evolution of the one-particle densities are governed by a generalized continuity equation. The latter may be derived by integrating the Smoluchowski equation, that is, the Fokker-Planck equation for a system of (colloidal) particles with {\em overdamped} stochastic equations of motion (i.e.\ the inertial terms in the microscopic Langevin equations of motion are neglected). The key approximation of the DDFT approach is that the non-equilibrium two-body density distribution functions at time $t$ are set equal to those of an {\em equilibrium} system with the same one-body density profile\cite{marconi:8032,marconi:a413, ArcherSpinDec, AR04}. As a consequence, the currents entering the DDFT equations are determined by (functional derivatives of) the {\em equilibrium} Helmholtz free energy functional. As in most DDFT applications so far, we neglect here the effect of hydrodynamic (solvent-induced) interactions between the particles.

\begin{figure*}[tpb]
\begin{center}
\includegraphics[width=2.\columnwidth]{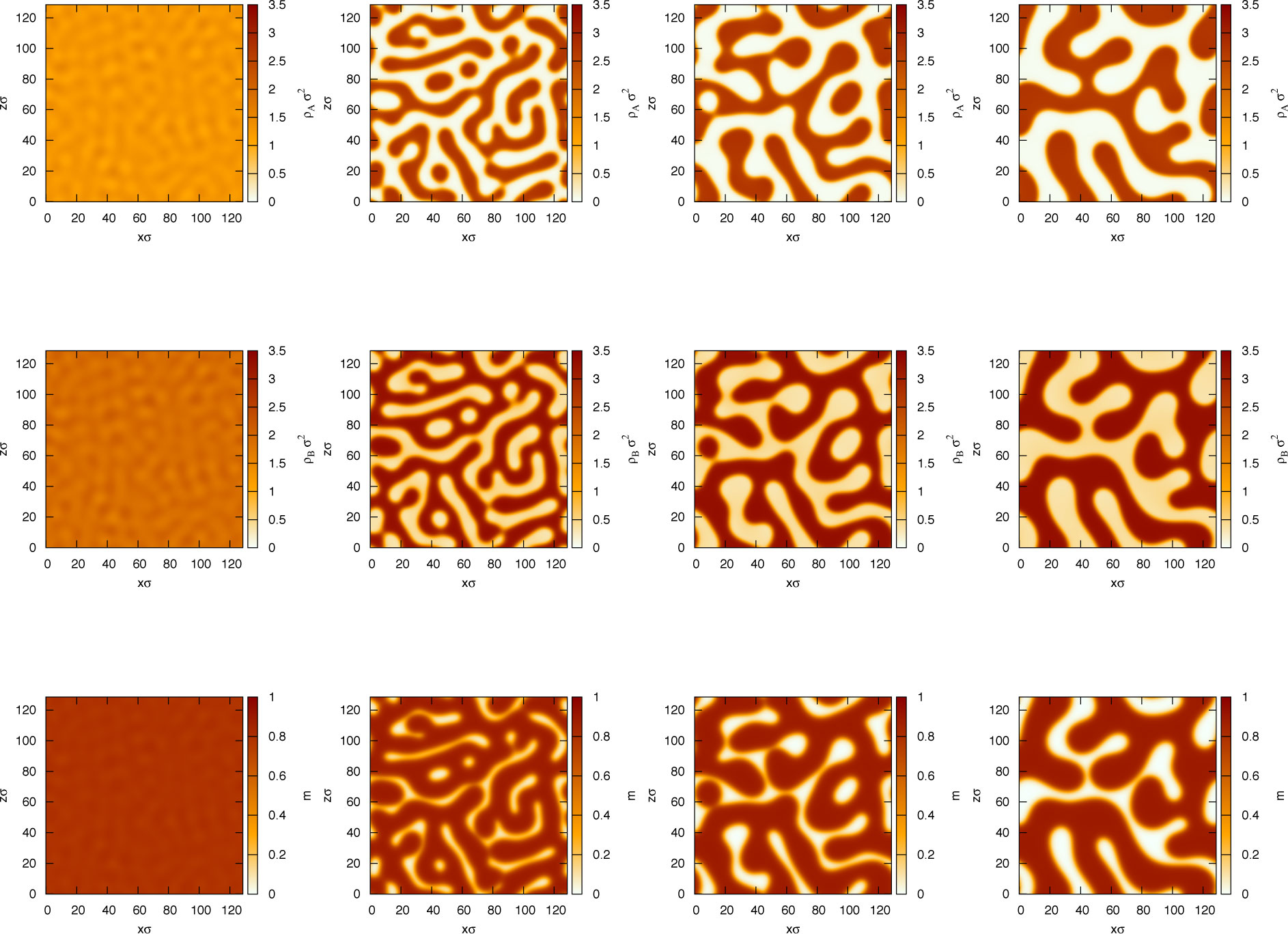}
\end{center}
\caption{Density profiles of the isotropic particles (upper row), magnetic particles (middle row) and the magnetization (bottom row) as a function of the position. The time increases from the left to the right: $t_1=60\tau_B$, $t_2=200\tau_B$, $t_3=400\tau_B$ and $t_4=800\tau_B$. The parameters are $\rho\sigma^2=3.2$, $x=0.6$, $\varepsilon^*=5.0$ and $\delta=0.1$.}
\label{fig.5}
\end{figure*}

For the present system, where one species (the $B$-particles) has internal degrees of freedom, one should employ the DDFT equations for anisotropic particles recently proposed in Ref.\ \onlinecite{RexLoewen}, which shows that the equations of motion for the position- and angle-dependent densities involve both the usual translational currents ${\bf j}_\alpha=-D_\alpha\rho_\alpha\nabla \delta \mathcal{F}/\delta \rho_\alpha$, and also `rotational current' terms resulting from application of the angular momentum operator to the thermodynamic driving force $\delta \mathcal{F}/\delta \rho_B$. However, the numerical solution of the resulting set of equations of motions for the demixing problems that we consider here (see below) involve simultaneously determining not only the two-dimensional (isotropic) number density profiles of the $A$- and $B$-particles, respectively, but also of the orientational distribution function $h_B({\bf r},\omega,t)$. The angle-dependence of the latter induces additional dimensions making the numerical calculations rather involved.

In the present study, we somewhat simplify the problem by assuming that the magnetic degrees of freedom are {\em at each moment in time} in equilibrium with the density profiles. Physically, this assumption implies that the relaxation time of the magnetic moments is much shorter than that of the translational degrees of freedom. This assumption implies that the functional derivative $\delta \mathcal{F}/\delta h({\bf r},\omega,t)=0$ at {\em all times} $t$, i.e., there is no driving torque. Under these conditions, the DDFT equations for the present system reduce to a coupled set of equations for the number density profiles of the $A$- and $B$- particles 
\begin{align}
 \Gamma_\alpha^{-1}\frac{\partial \rho_\alpha(\mathbf{r},t)}{\partial t}=\nabla\cdot \left[\rho_\alpha(\mathbf{r},t)\nabla \frac{\delta \mathcal{F}[\rho_A(\mathbf{r},t),\rho_B(\mathbf{r},t)]}{\delta \rho_\alpha(\mathbf{r},t)}\right],\label{eq.34}
\end{align}
combined with the self-consistency relation 
\begin{align}
 h_B(\mathbf{r},\omega,t)=\dfrac{\exp(\mathbf{B}(\mathbf{r},t)\cdot \mathbf{s}(\omega,t))}{\int d\omega \exp(\mathbf{B}(\mathbf{r},t)\cdot \mathbf{s}(\omega,t))},\label{ddftangle}
\end{align}
where the time-dependent effective field $\mathbf{B}(\mathbf{r},t)$ is given by Eq.\ \eqref{eq.15} and where $\mathcal{F}$ is the MF Helmholtz free energy functional developed in Sec.\ \ref{section:theDFT}. The mobility coefficients $\Gamma_\alpha$ in Eq.~(\ref{eq.34}) are related to the diffusion constants via $\Gamma_\alpha=D_\alpha/(k_BT)$, where $\alpha=A$ or $B$. In what follows we assume that these are equal: $\Gamma_A=\Gamma_B=\Gamma$.

When the external potential $V_{\rm ext}=0$, the uniform density distributions, $\rho_\alpha(\mathbf{r},t)=\rho_\alpha^{\rm bulk}$, always correspond to a stationary solution of the DDFT Eq.\ \eqref{eq.34}, since in this case the functional derivatives $\delta \mathcal{F}/\delta \rho_\alpha$ are constants. However, if one considers applying small harmonic perturbations to the uniform densities $\rho_\alpha(\mathbf{r},t)=\rho_\alpha^{\rm bulk}+\delta \rho_\alpha(\mathbf{r},t)$, where $\delta \rho_\alpha(\mathbf{r},t)\sim \sin(\mathbf{k}\cdot \mathbf{r})$, with wave number $|\mathbf{k}|=k$, and where $\delta\rho_\alpha\sigma^2 \ll 1$, then one finds that inside the spinodal region (the red region in Fig.~\ref{fig.1}) density fluctuations with certain wavenumbers $k$ grow with time\cite{ArcherSpinDec, PASTM11} -- i.e.\ within the spinodal the fluid is linearly unstable. Outside the spinodal, the system is linearly stable, i.e.\ the amplitude of any small amplitude density fluctuations decrease over time. In the region between the binodal and the spinodal (the black regions in Fig.~\ref{fig.1}), the fluid is linearly stable, but is not absolutely stable: if the amplitude of a given density perturbation is large enough, then the amplitude of this density fluctuation will grow over time. This is due to the non-linear terms in Eq.\ \eqref{eq.34}. For densities outside of the binodal, the uniform fluid is absolutely stable and all density modulations diminish in amplitude over time. The above description of the system is therefore qualitatively very similar to the results from Cahn-Hilliard theory\cite{CahnHilliard}.

Thus, there are two phase separation mechanisms: Firstly, spinodal demixing, which is triggered by the presence of small amplitude density modulations, which in reality are always present due to thermal fluctuations. This is the dominant mechanism inside the spinodal. Secondly, for state points in the region between the spinodal and the binodals, where the uniform fluid is linearly stable, phase separation must proceed via the nucleation of density fluctuations with sufficiently large amplitude. We present results pertaining to these two mechanisms below.

\subsection{Spinodal demixing}

To study the phase separation dynamics in the spinodal region, we set the time $t=0$ density profiles to be $\rho_\alpha(\mathbf{r},t=0)=\rho_\alpha^{\rm bulk}+\chi(\mathbf{r})$, where $\chi(\mathbf{r})$ is a small amplitude random white noise field, which is equivalent to adding many harmonic density perturbations, with randomly chosen amplitude, phase and wavenumbers $k$. The density and magnetization profiles are then evolved forward in time using Eqs.\ \eqref{eq.34} and \eqref{ddftangle}. Note that we only add noise to the initial $t=0$ density profiles and do not add noise at any other subsequent time. This corresponds to taking a uniform system and then rapidly quenching it into the unstable region of the phase diagram, by decreasing the temperature. Inside the spinodal, the density perturbations with wave numbers $0<k<k_c$ grow over time\cite{ArcherSpinDec, PASTM11}. The density modulations with wavenumber $k\approx k_*$, where $0<k_*<k_c$, grows fastest in amplitude over time, leading to density profiles having density modulations with a typical length scale $\approx 2\pi/k_*$, at short times after the quench. As is illustrated by the results displayed in Fig.\ \ref{fig.5}, the domains of demixed fluid then coarsen over time. Since in the present system this demixing is between a magnetic and a non-magnetic phase, we also observe a similar pattern in the local magnetization; see the bottom row in Fig.\ \ref{fig.5}. The results in this figure are for a fluid with total density $\rho\sigma^2=3.2$, concentration $x=0.6$ and with $\varepsilon^*=5.0$ and $\delta=0.1$. The phase diagram for this system is displayed in Fig.\ \ref{fig.1}. For a given total density $\rho$ within the spinodal region, on varying the concentration $x$ one may observe bicontinuous labyrinthine patterns, such as those displayed in Fig.\ \ref{fig.5}, or alternatively one observes phase separated morphologies consisting of `islands' of the minority phase surrounded by a `sea' of the majority phase. A more detailed discussion of spinodal phase separation and the resulting structures that we observe in the present system will be published elsewhere. 

\subsection{Nucleation}
\label{sec:Nucleation}

Before using DDFT to investigate the dynamics of nucleation, it is worth recalling the main results from classical nucleation theory (CNT) and also approaches to nucleation using equilibrium DFT.

\subsubsection{\label{sec:CNT}Classical nucleation theory}

Nucleation is normally considered to be the relevant phase separation mechanism within the {\it metastable} regions of the phase diagram, i.e., within the regions between the binodals and the spinodal. In these metastable regions, one may consider forming (circular, in 2D) clusters of the new (globally stable) phase with radius $R$, surrounded by the metastable bulk phase. One finds that the free energy as a function of $R$ initially increases, reaches a maximum at $R=R_c$, the critical radius, and then decreases for $R>R_c$. It is assumed that clusters of a given radius are randomly formed in the system by thermal fluctuations. Clusters with radius $R<R_c$ then typically shrink, since this lowers the system free energy. On the other hand, clusters with $R>R_c$ must grow without limit (in an infinite size system), since doing this also reduces the system free energy, thereby initiating the transformation into the new (stable) phase. Thus, the excess free energy corresponding to the `critical cluster', i.e.\ the cluster with radius $R_c$, is important, because this is the free energy barrier which must be surmounted for the phase separation to occur.

CNT treats the nucleation process on the simplest possible level. The key assumption of CNT is that {\em any} cluster (regardless of its actual size) can be regarded as a macroscopic object with a homogeneous density (and thus, pressure) inside and outside its surface. Moreover, interfacial curvature effects are typically neglected. As a consequence, the free energy related to creation of a nucleus can be written as a sum of two terms: a negative contribution stemming from the difference of the (bulk) pressures inside and outside the nucleus, and a positive contribution related to the increase of surface free energy. The latter is determined by the interfacial tension $\gamma$ of a planar interface (we use the interfacial tension calculated in Sec.\ \ref{sec:interface_structure} above). Applying this concept to the two-dimensional system at hand, the grand potential for the creation of an $A$-rich phase nucleus in the surrounding ``sea" of $B$-rich phase is
\begin{align}
\Omega^\text{CNT}(R)=-P_B{\cal A}-\pi R^2|\Delta P|+2\pi R\gamma
\label{eq.35},
\end{align}
where ${\cal A}$ is the total system area, $\Omega_0\equiv -P_B{\cal A}$ is the grand potential of the uniform $B$-rich phase and $\Delta P$ is the difference between the bulk pressure $P_B$ and the pressure of the $A$-rich phase that is being nucleated. The nucleation barrier is then given by the maximum of $\Delta\Omega^\text{CNT}=\Omega^\text{CNT}-\Omega_0$ which follows from Eq.~(\ref{eq.35}) as
\begin{align}
\Delta\Omega_c^\text{CNT}=\dfrac{\pi\gamma^2}{|\Delta P|}.\label{eq.barrier}
\end{align}
The corresponding critical radius is 
\begin{align}
R_c=\dfrac{\gamma}{|\Delta P|}\label{eq.36}.
\end{align}
For $R>R_c$ the grand potential decreases, indicating the absence of a sustaining force against the growth of a drop of the new phase.

\begin{figure}[tpb]
\includegraphics[width=8cm]{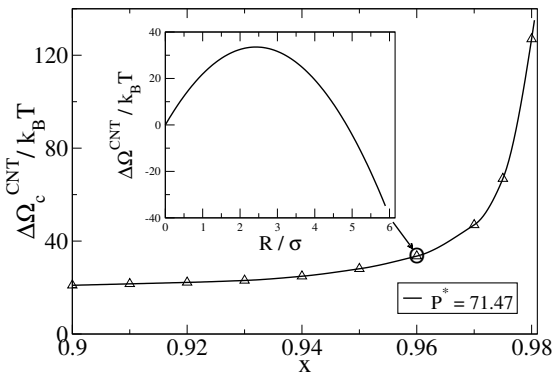}
\caption{The nucleation barrier $\Delta \Omega_c^\text{CNT}$ obtained from classical nucleation theory as a function of the concentration $x$ for fixed total density $\rho\sigma^2=3.2$. The inset shows the excess grand potential of the nucleus as a function of the radius $R$ for fixed concentration $x=0.96$. The coupling parameters are $\varepsilon^*=5.0$ and $\delta=0.1$.}
\label{fig.6}
\end{figure}
\begin{figure}[tpb]
\includegraphics[width=8cm]{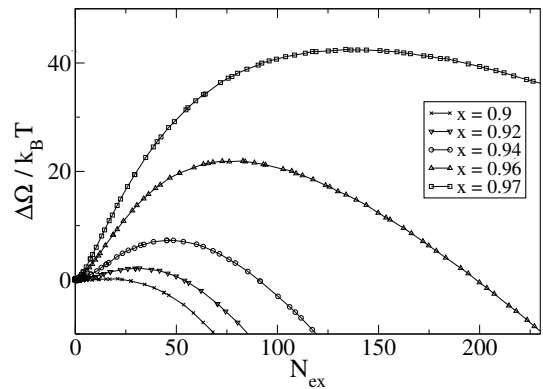}
\caption{The grand potential $\Delta \Omega$ of the nucleus obtained from the DFT approach as function of the excess number $N_\text{ex}$ for different concentrations. The parameters are $\rho\sigma^2=3.2$, $\varepsilon^*=5.0$ and $\delta=0.1$.}
\label{fig.7}
\end{figure}

In the present calculations we focus on nucleation on the right-hand side of the phase diagram, i.e., the formation of non-magnetic clusters predominantly containing $A$-particles, out of the ferromagnetic, $B$-dominated liquid. We consider the nucleus and its environment (i.e., the undersaturated magnetic liquid) at the same pair of chemical potentials $\mu_A$, $\mu_B$. Because we are working on the right side of the phase diagram, these chemical potentials are typically smaller than those corresponding to phase coexistence. The quantity $\Delta P$ appearing in Eqs.~(\ref{eq.35}), (\ref{eq.barrier}) and (\ref{eq.36}) is then defined as the difference between the pressure corresponding to the actual (magnetic) state on the right side, and the pressure of the corresponding nonmagnetic state on the left hand side of the phase diagram. These states with equal chemical potentials (as well as the associated pressures) are found from the bulk free energy given in Eq.~\eqref{eq.18}, which yields the desired quantities through the relations
\begin{align}
\mu_A&=f+\rho\left( \frac{\partial f}{\partial \rho}\right)_x - x\left( \frac{\partial f}{\partial x}\right)_\rho \\
\mu_B&=f+\rho\left( \frac{\partial f}{\partial \rho}\right)_x +(1- x)\left( \frac{\partial f}{\partial x}\right)_\rho \\
P&=\rho^2\left(\frac{\partial f}{\partial \rho}\right)_x
\label{eq.thermo_relations}
\end{align}
In Fig.\ \ref{fig.6} we plot CNT results for the free energy barrier height $\Delta \Omega_c^\text{CNT}$ for a range of concentrations $x$ of the magnetic species within the metastable regime. The chosen path corresponds to states at constant total density $\rho\sigma^2=3.2$. The inset shows the grand potential [see Eq.~(\ref{eq.35})] as a function of $R$ for one particular state point. As expected from the structure of Eq.~(\ref{eq.barrier}), the nucleation barrier according to CNT becomes infinitely large at the coexistence line, where $\Delta P=0$. For the total density $\rho\sigma^2=3.2$, the binodal is at the concentration $x=0.985$. Decreasing then the concentration towards the spinodal value, the barrier height decreases. Note, however, that the barrier directly at the spinodal is not exactly zero, as one would expect at the limit of metastability. This deficiency is a well-known artefact of CNT. Indeed, given that CNT is a macroscopic theory, it is not surprising that its predictions become unreliable when the critical clusters become so small that they contain only a few particles. Under such conditions, a microscopic theory such as DFT is clearly more appropriate.

\subsubsection{DFT approach to nucleation}

Since the important work of Oxtoby and Evans\cite{oxtoby:7521} there has been much work done using DFT to obtain a more reliable (microscopic) estimate for the free energy barrier $\Delta\Omega_c$ -- see e.g.\ Refs.\ \cite{talanquer:5190, UC07, LutskoEPL, Lutsko_JCP_2009_1, ArcherEvansNucleation} for examples of recent work. Here, we use the Oxtoby-Evans approach to study the nucleation of isotropic clusters of the phase rich in $A$-particles, and compare the resulting free energy barrier with the corresponding CNT results discussed in Sec.\ \ref{sec:CNT}.

\begin{figure}[tpb]
\includegraphics[width=7.5cm]{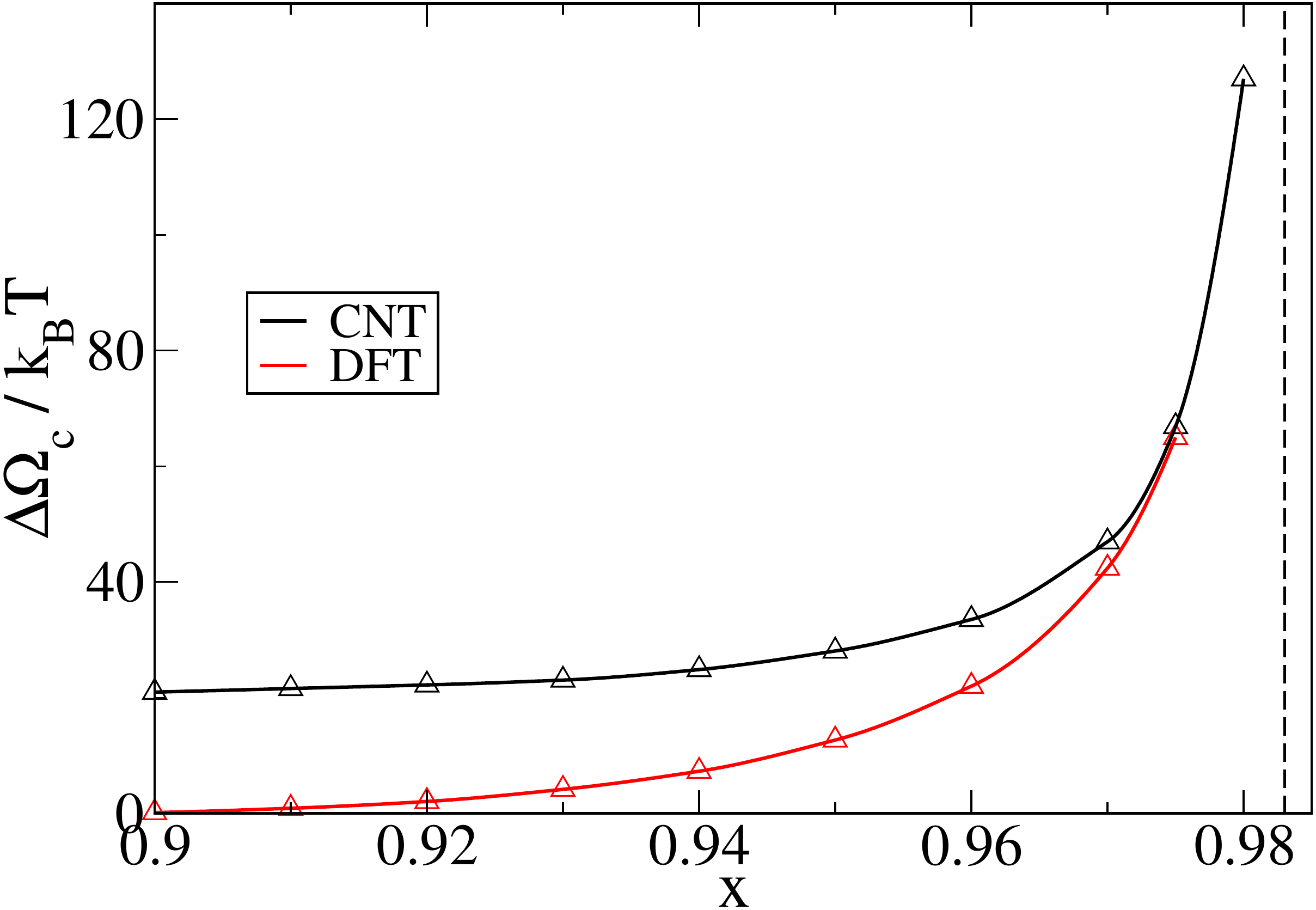}
\caption{The nucleation barrier $\Delta \Omega_c$ obtained from classical nucleation theory (black curve) and from the DFT approach (red curve) as function of the concentration $x$. The dashed curve denotes the state point of coexistence. The parameters are $\rho\sigma^2=3.2$, $\varepsilon^*=5.0$ and $\delta=0.1$.}
\label{fig.8}
\end{figure}

The key idea in all equilibrium DFT based approaches to nucleation is that the density profile characterizing the critical nucleus corresponds to a {\em saddle point} of the grand canonical free energy\cite{oxtoby:7521}. Assuming a symmetric droplet in the center of the system, the goal is thus to find the density profiles $\rho_\alpha({\bf r},\omega)=\rho_\alpha(x,z,\omega)$ for which
\begin{equation}
\frac{\delta\Omega[\{\rho_\alpha\}]}{\delta \rho_\alpha({\bf r},\omega)}=\frac{\delta \mathcal{F}[\{\rho_\alpha\}]}{\delta \rho_\alpha({\bf r},\omega)}-\mu_\alpha=0,
\end{equation}
with the boundary conditions
\begin{align}
\lim\limits_{r\to\infty}\rho_\alpha(x,z,\omega)=&\left.\rho_\alpha^{\text{bulk}}(x,z,\omega)\right\vert_{\{\mu_\alpha\}}\\
=&\rho_\alpha^{\text{bulk}}h_\alpha^{\text{bulk}}(\omega),
\end{align}
where $r=\sqrt{x^2+z^2}$. As demonstrated by Oxtoby and Evans\cite{oxtoby:7521}, these density profiles may be found by iterating the Euler-Lagrange equations, beginning with a simple approximation for the initial profiles. The latter are characterized by spherical symmetry and a sharp (step-wise) change of the number density at a radius $R$. The idea then is that if the guessed radius is too small (large), the droplet will shrink (grow) during the iteration procedure until the profiles eventually approach the density values corresponding to the globally stable (unstable) phase. However, to identify the critical droplet one must iterate the Euler-Lagrange equations a limited number of times (for each initial guess $R$). This procedure allows for an estimate of the grand potential $\Omega$ as function of $R$. The critical droplet then follows as the position of the maximum.  Inspired by these ideas we have performed additional calculations, not with the DDFT (which will be discussed below), but with a simple, relaxational algorithm which is equivalent to the Oxtoby-Evans method. Importantly, this algorithm keeps the chemical potentials fixed (just as in the original Oxtoby-Evans work\cite{oxtoby:7521}), while the number densities themselves are not conserved. Using this algorithm we investigated the evolution of several initial
profiles of the form 
\begin{align}
 \rho_\alpha(x,z)&=\begin{cases}
  \rho_\alpha^{\text{I}},  & \text{if}\ \ r<R\\
  \rho_\alpha^{\text{II}}, & \text{else}
\end{cases},\nonumber\\
m(x,z)&=\begin{cases}
  0,  & \text{if}\ \ r<R\\
  m^{\text{II}}, & \text{else}
\end{cases}.
\end{align}
The initial values for the densities inside ($\rho_\alpha^{\text{I}}$) and outside ($\rho_\alpha^{\text{II}}$) the nucleus, are set to the bulk densities determined by the chosen value of the chemical potentials. The magnetization for these state points is obtained self-consistently from Eq.~\eqref{eq.15}. For each initial guess, we performed about 300 iterations of the density profiles and the corresponding orientational profile, keeping the chemical potentials fixed. In this way we obtained an estimate of the excess free energy $\Delta \Omega$ for each given cluster size. As it turns out, the actual density profiles characterizing the clusters are rather smooth, such that the definition of a radius becomes ambiguous. As an alternative `reaction coordinate', we thus consider the quantity
\begin{align}
 N_\text{ex}=\int\negthickspace dx\negthickspace\int\negthickspace  dz \left(\rho_A(x,z)-\rho^\text{bulk}_A\right).
\end{align}
which counts the number of $A$-particles in the cluster (recall that we are considering the nucleation of isotropic $A$-dominated clusters). Numerical results for the functions $\Delta\Omega(N_\text{ex})$ at various chemical potentials within the metastable regime are plotted in Fig.~\ref{fig.7}. More precisely, the chemical potentials considered are associated to concentrations (of $B$-particles) between the spinodal and the binodal along a path with fixed total number density $\rho\sigma^2=3.2$ (CNT results along this path are presented in Fig.\ \ref{fig.6}). As may be seen from Fig.\ \ref{fig.7}, all the curves reveal a clear maximum and thus, a clearly identifiable nucleation barrier, the height and position of which increase upon increasing the associated (bulk) concentration $x$. This finding is fully consistent with our expectation that the nucleation barrier and the size of the critical droplet are smallest close to the spinodal and then increase monotonically upon approaching coexistence. From the positions of the maxima in $\Delta\Omega(N_\text{ex})$ we also see that typical critical droplets contain between a few ten and a few hundred of particles, consistent with results of other DFT studies\cite{ArcherEvansNucleation,Lutsko_JCP_2009_1}. Our data for the height of the nucleation barrier as function of the concentration are summarized in Fig.\ \ref{fig.8}, where we have included the corresponding macroscopic (CNT) results from Fig.\ \ref{fig.6}. The main difference between the two approaches is that the microscopic DFT calculation yields, contrary to CNT, a {\em vanishing} nucleation barrier at the spinodal, as one should expect on physical grounds. On the other hand, approaching the binodal the two curves merge, reflecting the increasingly macroscopic character of the critical cluster.

\begin{figure}[tpb]
\includegraphics[width=7.5cm]{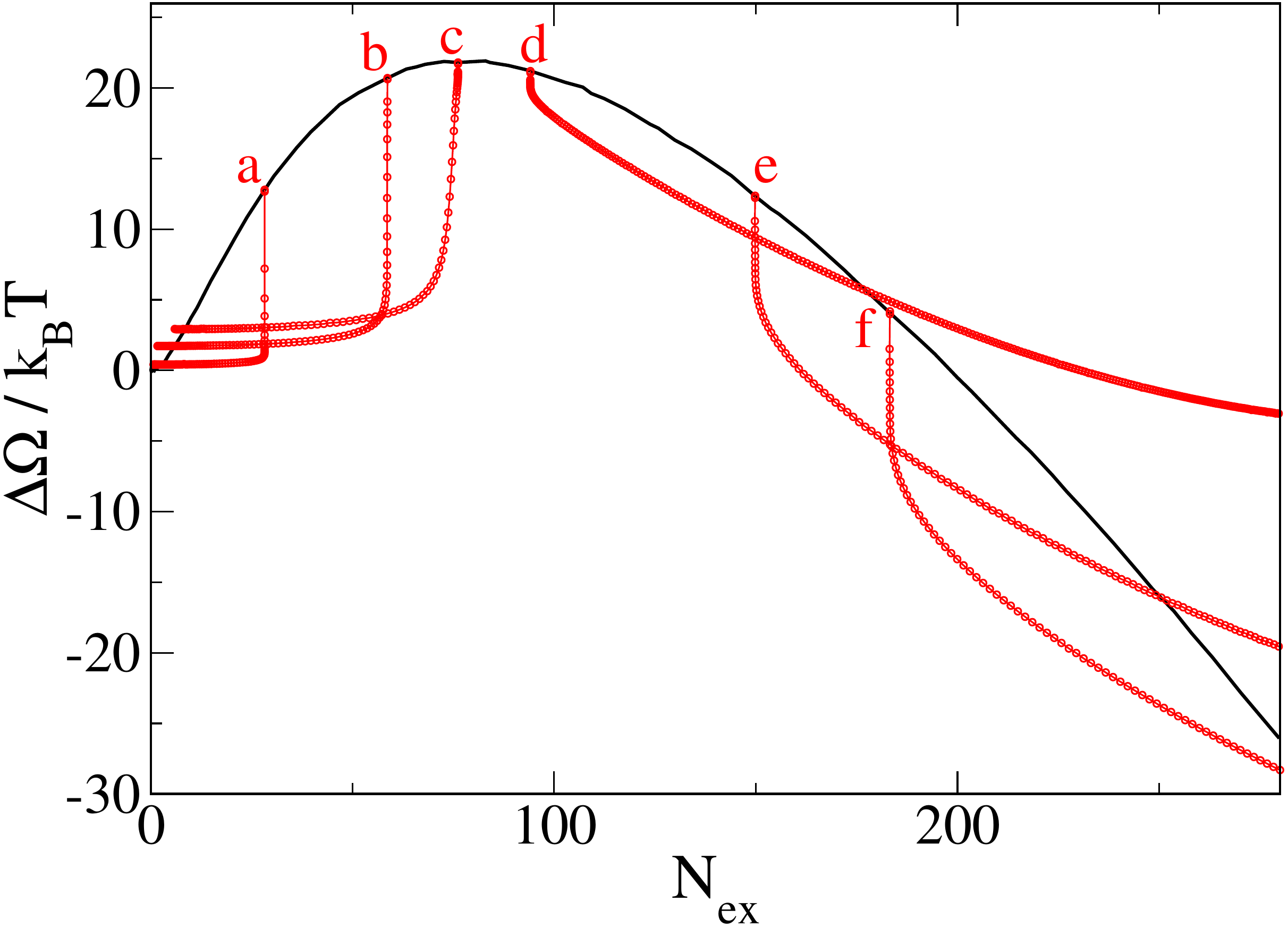}
\caption{The grand potential $\Delta \Omega$ of the nucleus as function of the excess number $N_\text{ex}$. The black curve is obtained from the DFT approach and the red curve corresponds to DDFT calculations for initial configurations with radius $R=2.05\sigma$ (a), $R=2.5\sigma$ (b), $R=2.75\sigma$ (c), $R=3\sigma$ (d), $R=3.6\sigma$ (e), $R=4\sigma$ (f). The parameters are $\rho\sigma^2=3.2$, $x=0.96$, $\varepsilon^*=5.0$ and $\delta=0.1$.}
\label{fig.9}
\end{figure}

\subsubsection{DDFT approach to nucleation}

\begin{figure*}[tpb]
\begin{center}
\includegraphics[width=2.\columnwidth]{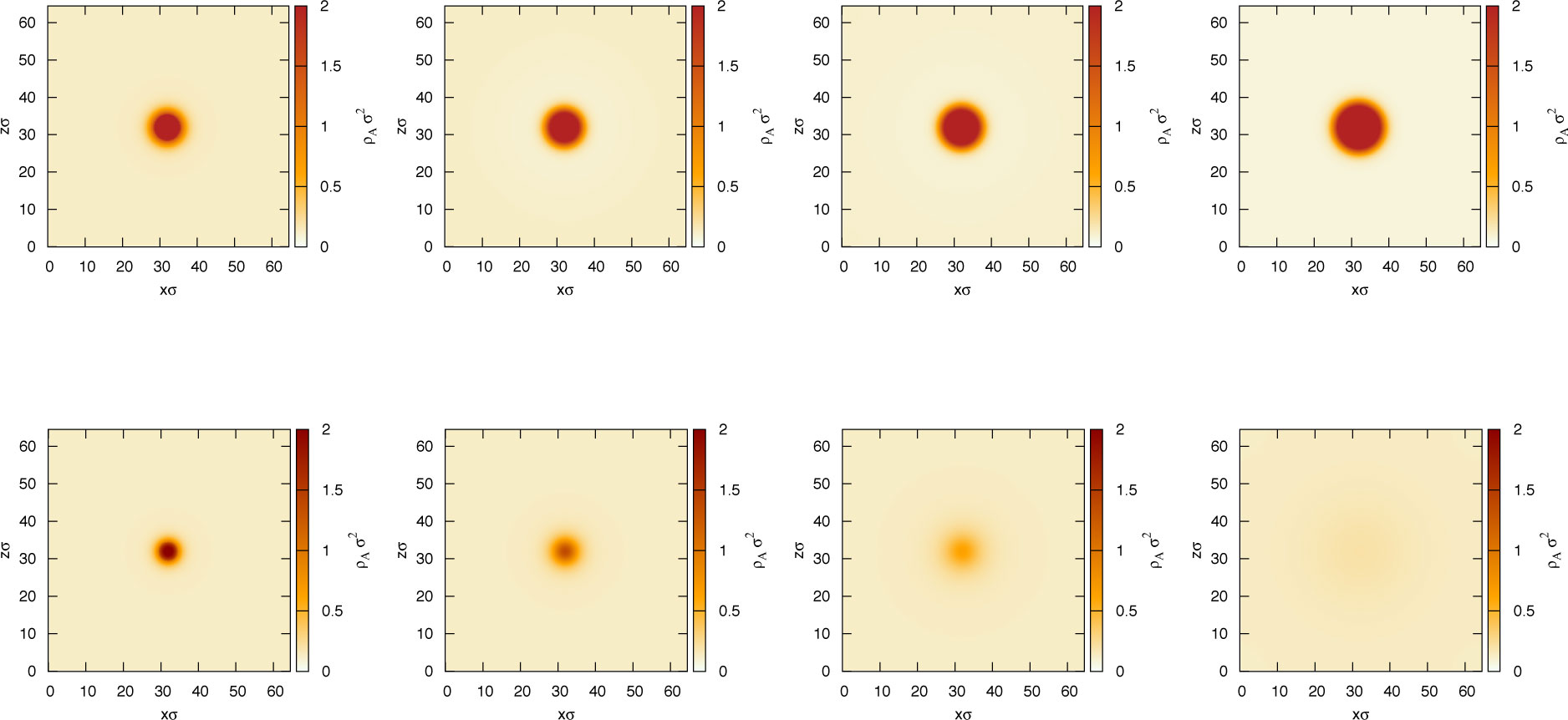}
\end{center}
\caption{Density profiles of the isotropic particles (species $A$) as function of the position. The upper row shows a supercritical growing nucleus ($R^\text{initial}=3.6\sigma$) for subsequent times. The bottom row shows a time sequence for a subcritical nucleus ($R^\text{initial}=2.5\sigma$). The time increases from the left to the right. Upper row: $t_1=0\tau_B$, $t_2=200\tau_B$, $t_3=600\tau_B$ and $t_4=2000\tau_B$. Bottom row: $t_1=0\tau_B$, $t_2=40\tau_B$, $t_3=80\tau_B$ and $t_4=200\tau_B$. The parameters are $\rho\sigma^2=3.2$, $x=0.96$, $\varepsilon^*=5.0$ and $\delta=0.1$.}
\label{fig.10}
\end{figure*}

Whilst it is clear that the above and other equilibrium DFT based approaches are able to calculate the density profiles corresponding to the critical droplet, it is not clear whether the other density profiles (i.e.\ those not corresponding to the critical droplet) have any physical significance. Of particular interest are the density profiles corresponding to the most likely pathway (MLP) that go up to and then descend from the critical droplet state\cite{LutskoEPL, Lutsko_JCP_2009_1, ArcherEvansNucleation, Lutsko/arXiv}. These MLP profiles should correspond to what one would observe experimentally for nucleation in the system.

Recently, Lutsko\cite{Lutsko/arXiv} argued that one should be able to determine the MLP using DDFT by initiating the system at the critical droplet density profiles. Since these profiles correspond to a saddle point, and in reality one is never able to initiate the system {\em exactly} at the saddle point, the density profiles evolve under the DDFT away from this point. There are two directions the system may move: firstly, towards a state corresponding to the drop disappearing, with the excess density being spread uniformly throughout the system and secondly, the drop may grow.

The performance of DDFT in this context is not yet fully understood (see Ref.~\onlinecite{Lutsko/arXiv}), contrary to with the static DFT approaches mentioned above. One issue in this context is the fact that, within the DDFT, the number densities are {\em conserved} quantities [as reflected by the appearance of a divergence in front of the free energy derivative in Eq.~(\ref{eq.34})]. This is in contrast to traditional (static) DFT approaches towards nucleation, where the fixed quantity is the chemical potential(s). In view of this subtle point, and given the rather plausible results from the DFT approach described so far, it is an important question whether a different algorithm, and particularly the conserved dynamics implied by the DDFT [see Eqs.(\ref{eq.34})-(\ref{ddftangle})], yields consistent results. We recall that the DDFT is constructed such that the density evolves towards a profile which minimizes the free energy (one can prove that the free energy always decreases or remains constant under the time evolution of the DDFT, unless the system is externally driven\cite{PASTM11}). Therefore, one would expect that the critical profile found in the approach discussed above, also `behaves' as a saddle point within DDFT calculations. To confirm this, we have performed a number of DDFT calculations with initial density profiles stemming from the DFT calculations described above (after 300 iterations). Some results of these calculations are illustrated in Fig. \ref{fig.9}. Note that for the DDFT results we have used a slightly different definition of $N_\text{ex}$. The DDFT calculations are performed on a finite size square area of length $L=64\sigma$, with periodic boundary conditions. We start with the drop located at the center and we define
\begin{align}
 N_\text{ex}=\int\negthickspace dx\negthickspace\int\negthickspace  dz \left(\rho_A(x,z)-\rho^\text{corner}_A\right),
\end{align}
where $\rho^\text{corner}_A=\rho_A(0,0)= \rho_A(0,L)=\rho_A(L,0)=\rho_A(L,L)$ is the value of the density at the corners of the (square, periodic) system. During the initial stages of the evolution $\rho^\text{corner}_A=\rho_A^\text{bulk}$, and so the value of $N_\text{ex}$ remains constant due to the conserved dynamics. However, at later times $\rho^\text{corner}_A$ changes, and so $N_\text{ex}$ changes with time. This is because either excess density from the center of the system diffuses out to the corner as the drop disappears, or because as the drop grows it removes particles of species $A$ from the surrounding fluid and so $\rho^\text{corner}_A$ decreases. In Fig.\ \ref{fig.9} the curves labelled by a,b,c correspond to calculations where the radius characterizing the initial profile is smaller than that characterizing the critical droplet predicted by the previous DFT approach. On the other hand, the curves labelled by d,e,f have been started from ``supercritical" clusters. In all cases, the DDFT algorithm evolves in the direction predicted by the previous free-energy approach. That is, when starting from a subcritical or supercritical profile, respectively, the droplet vanishes or grows without restriction. We also see from Fig.\ \ref{fig.9} that the actual values of $\Delta\Omega$ ``on the way" towards the final state are strongly different from those predicted by the previous approach. The interpretation of this issue clearly needs further investigation (see also the discussion in Ref.~\onlinecite{Lutsko/arXiv}). Furthermore, in the limit $N_\text{ex}\rightarrow 0$ the DDFT curves a,b,c approach different values, which depend on the initial profile. This is a consequence of the fact that the initial profiles a-c correspond to different space-averaged densities. This stems from the fact that the DDFT conserves the total densities, and so the final state emerging for subcritical clusters does not necessarily have chemical potentials equal to those chosen in the previous DFT calculations. We conclude that, at least, the DDFT predictions for the critical nucleus are consistent with the traditional DFT theory.  Typical density profiles illustrating the nucleation dynamics according to DDFT on a microscopic (space-resolved) level are shown in Fig.\ \ref{fig.10}. Besides the shrinking/growth process, we remark in particular the diffuse character of the density profiles, which directly reflects the difficulty in associating a fixed radius to the instantaneous droplets. 

\section{Concluding remarks}
In this paper we have investigated the demixing phase transition of a binary mixture of magnetic and non-magnetic soft-core particles from both a static and a dynamic point of view. To this end we have employed
classical equilibrium DFT as well as the recently developed DDFT\cite{marconi:8032,marconi:a413, ArcherSpinDec, AR04}, a time-dependent extension of DFT involving a generalized continuity equation for the density.

Our analysis of the equilibrium phase behavior of the system shows that the magnetic (Heisenberg) interaction within one species is capable of inducing macroscopic phase separation. More precisely, for sufficiently large ferromagnetic coupling strength we find
a combined phase transition where the system both demixes and develops global ferromagnetic ordering in one of the phases. Depending on the total density, the transition may either be second or first order (in terms of the magnetization and composition, respectively), with the two regimes being separated by a tricritical point. 
Thus, the general topology of the phase diagram (in the fluid regime), particularly the appearance of a Curie line and a tricritical point, coincides with that of 3D Heisenberg and XY-fluids\cite{Omelyan,PhysRevE.68.061510}. Based on DFT we have also calculated the structure of the fluid-fluid interface in the first-order regime, i.e.\ we calculate the density and magnetization profiles, as well as the resulting interfacial (line) tension. As expected, the latter vanishes upon approaching the tricritical point from the high-density side.

We note that the present DFT results are based on a mean-field (MF) approximation for the excess free energy contribution from both the soft-core repulsion (where MF theory has already proven to be very accurate\cite{PhysRevE.62.7961}) and from the Heisenberg interactions. Thus, the quantitative reliability of our results remain to be checked against those of more refined free energy functionals and/or computer simulations. However, judging from previous theoretical studies of 3D Heisenberg fluids\cite{PhysRevE.58.3426} we would expect the MF results to be quite good. 

The second part of the paper has been devoted to the dynamics of the (first-order) phase separation, concentrating mainly on the nucleation of non-magnetic clusters within the metastable ferromagnetic phase. We have analyzed and compared results from three different approaches to nucleation. The simplest one is CNT, a macroscopic approach, involving the bulk pressure and line tension. The resulting nucleation barriers show familiar behavior (as compared to other model systems\cite{ArcherEvansNucleation,Lutsko_JCP_2009_1}) and predicts that the barrier height increases as one approaches to the binodal. In this regime CNT is expected to give a good approximation for $\Delta\Omega_c$. However, CNT incorrectly predicts that the barrier is finite at the spinodal. As a second step we have investigated the nucleation using the microscopic DFT approach originally proposed by Oxtoby and Evans\cite{oxtoby:7521}. In this approach one calculates the free energy as function of an appropriate ``reaction coordinate", which we set to the excess number of particles in the nucleus. The results for the nucleation barrier are very similar to those from CNT when the fluid state point is near to the binodal. However, in contrast to CNT, the DFT predicts that the nucleation barriers vanishes upon approaching the spinodal, as one should expect on physical grounds.

Finally, we have touched the issue of the nucleation pathway by comparing the results of the microscopic DFT with those from DDFT. Since both of these approaches are based on the same free energy functional, they are both probing the same underlying free energy landscape. The choice of reaction coordinate to some extent determines what regions of this landscape are accessed and what path to and from the saddle point are predicted. In addition, the `dynamics' inherent in each approach also determines this path. The DDFT has a real (physical) dynamics, which conserves the total number of particles in the system, whilst in contrast the Oxtoby--Evans DFT approach, which has a finite number of iterations starting from an initial guess, leads to having an effective (unphysical) dynamics, which does not conserve the number of particles. Thus, the DDFT is essentially a canonical theory, whereas the DFT approach is more grand canonical in character. As we have shown, these subtleties are irrelevant for the actual height of the nucleation barrier, but yield marked differences when we consider the evolution of profiles away from that of the critical nucleus towards equilibrium. To our knowledge, these two DFT approaches have not been directly compared before. What is clear is that more investigations must be done to understand in detail the role of DDFT for the nucleation pathway, and more importantly which approach is the relevant one for determining the true MLP.

There are several directions which we believe require further investigation.
First, as a direct extension of the present work, it would be interesting to consider in more detail the spinodal decomposition occurring in the 
unstable range of the phase diagram. Our results in the present study already indicate that the DDFT is capable of describing the coarsening process characterizing this regime in a qualitative way (see Sec.\ IV A). However, this issue clearly calls for a more systematic investigation including a (linear) stability analysis\cite{ArcherSpinDec} and also an investigation of dynamic correlations such as the time-dependent structure factor. This quantity may be obtained using recent extensions of the DDFT to determine dynamical two-particle correlation functions\cite{PhysRevE.75.040501}. 

Another (maybe more technical) open point concerns the effect of the actual formulation of the DDFT. In the present study we have restricted ourselves to a simplified version where the spin degrees of freedom relax instantaneously; therefore, we were essentially dealing with the DDFT equations for isotropic particles, and the magnetization entered only via the selfconsistency relation for the (local) magnetization. However, to further proceed one should also explore results from the full DDFT for anisotropic particles, which involves an equation for the rotational dynamics on top of that of the translational dynamics\cite{RexLoewen}. A particular interesting question in this context concerns the role of different time scales of rotational and translational motion for the demixing dynamics.

Finally, it should be interesting to extend the present study towards 2D (magnetic) systems exposed to {\em patterned} surfaces and, optionally, additional external magnetic fields. In fact, a number of recent experiments\cite{PhysRevLett.100.148304,PhysRevLett.99.038303} have addressed the question of phase behavior and transport of magnetic particles on complex substrates\cite{FischerGarnet11}. All these question can be, in principle, tackled by the present (D)DFT approach. Work in these directions is in progress.

\acknowledgements

AJA gratefully acknowledges RCUK for support. KL and SHLK acknowledge financial support via the Collaborative Research Center (SFB) 910 ``Control of self-organizing nonlinear systems: Theoretical methods and concepts of application" and the Research Training Group (GRK) 1558
``Nonequilibrium Collective Dynamics in Condensed Matter and Biological Systems".

%

\end{document}